\documentstyle[psfig]{l-aa}     

\def \B {\mbox{B}}
\def \Bj {\mbox{B}_J}
\def \R {\mbox{R}}
\def \Rf {\mbox{R}_F}
\def \V {\mbox{V}}
\def \rms {{\it rms}}
\def \mic {~{$\mu$m}}
\def \etal {{et~al.\ }}

\begin{document}
 
   \thesaurus{3(11.05.2;11.06.2;11.12.2;11.16.1;11.19.7)}
   \title{Galaxy evolution at low redshift? -- I. Optical counts\thanks{Based in part on observations made
   at the CERGA Schmidt Telescope (Observatoire de la c\^ote d'Azur, France), at Observatoire de Haute-Provence
   (France) and at the European Southern Observatory (La Silla, Chile). ESO red survey, SERC blue survey
   and CERGA plates were digitized with MAMA (Machine Automatique \`a Mesurer pour l'Astronomie) which
   is developed and operated by INSU/CNRS.}}
   \author{E.~Bertin\inst{1,2} and M.~Dennefeld\inst{1,3}}
   \offprints{E.~Bertin (bertin@iap.fr)}
   \institute{Institut d'Astrophysique de Paris, 98bis Boulevard Arago, F-75014 Paris, France
        \and  European Southern Observatory, Casilla 19001, Santiago 19, Chile
	\and  Universit\'e Pierre \& Marie Curie, 4, place Jussieu, F-75005 Paris, France}
   \date{Received 22 December 1995; accepted 2 February 1996}
   \maketitle 
   \begin{abstract}
%
   We present bright galaxy number counts in the blue ($16<\Bj<21$) and red
   ($15<\R<19.5$) passbands, performed over 145 sq. degrees, both in the northern and
   southern galactic hemispheres.  The work was conducted on Schmidt plates
   digitized with the MAMA machine, and individually calibrated using an adequate
   number of CCD sequences. We find a relatively large density of bright galaxies
   implying a ``high'' normalization of the local luminosity function.
   Our counts and colour distributions exhibit no large departure from what
   standard no-evolution models predict to magnitudes $\Bj<21$, removing the need for
   evolution of the non-dwarf galaxy population in the optical, out to $z \approx 0.2$.
   This result disagrees with that of Maddox et al. (1990) on the APM catalog.
   We show that the APM and similar catalogs may be affected by a systematic magnitude
   scale error which would explain this discrepancy.

   \end{abstract}
%
%

\section{Introduction}
  Since Hubble (\cite{hubble}), galaxy number counts have been widely used as a statistical
  tool for probing the distant universe, with the hope of constraining both its geometry and the evolution of its content.
  Modern, high efficiency instruments and detectors
  reach impressive magnitude limits in the optical: $\Bj \approx 28$
  (Tyson \cite{tyson}, Metcalfe et~al. \cite{metcalfe:alb}, Smail et~al. \cite{smail:al}), although on a limited area.
  Galaxy counts done at brighter magnitudes ($\Bj<21$) are equally useful; their interpretation is much less
  model-dependant because of the smaller lookback-times and weaker cosmological effects. They thus constitute the
  link between models and deeper counts by providing a normalisation of both space densities and colours at low redshift.

  The problem with bright galaxy counts is that they obviously require substantial solid angles to be surveyed in order
  to provide
  statistically significant samples. Until very recently, such
  areas could only be surveyed in a reasonable time using photographic Schmidt plates. Since the mid-eighties,
  fast microdensitometers coupled with image analysis computer programs have been employed to produce automatically
  highly complete catalogs from large areas of the high galactic latitude sky like the COSMOS (Heydon-Dumbleton et~al.
  \cite{heydon:al}), the MRSP (Seitter \cite{seitter}), or the APM (Maddox et~al. \cite{maddox:alb}) galaxy surveys.

  The main difficulty when dealing with photographic material concerns flux measurements. Many scientific issues typically
  require the systematic errors to be kept $\la 0.1$~mag.. This is quite a difficult task to achieve on large scales, and
  can be reached only if a large number of {\em galaxy} standards per Schmidt plate, over the whole magnitude range of the counts,
  are observed to calibrate the data (Metcalfe et~al. \cite{metcalfe:alc}).
  Having such a high density of photometric standards is practically not possible with very large Schmidt plate surveys, 
  because of the huge observing time it would require. The catalogs extracted from these surveys are therefore likely to be
  hampered by systematic errors in their photometry, rather than by limited statistics.

  In this paper we re-examine galaxy number counts in the blue and red photographic passbands over the magnitude range
  $16<\Bj<21$ and $15<\Rf<19.5$. Our surveyed area is modest (140 sq. deg.), but has been carefully calibrated using
  a fair density of CCD standards ($\approx 700$, half of them beeing galaxies), as an attempt to keep photometric systematic
  errors $\la 0.1$~mag. {\em over the whole magnitude domain} of the survey. The procedure is described in detail in
  \S \ref{par:dataproc}, as well as the general data processing. The galaxy number counts are derived and compared to previous
  studies in \S \ref{par:numbercounts}. Model predictions about number counts and galaxy colour distributions are tested in
   \S \ref{par:modcomp}. We finally discuss the implications of our results in \S \ref{par:discuss}.

\section{The data}
  Our optical catalog was originally built as part of a program of identification of
  faint IRAS sources on Schmidt plates (Bertin \etal, in preparation) near both ecliptic poles.
  The 7 selected fields are characterized by
  a very low infrared cirrus emission, which ensures that the interstellar extinction 
  is negligible in the visible. Their coordinates are given in Table~\ref{tab:fields}.
  To each field corresponds one plate in the blue photographic passband, and one in the red.
  Northern plates were taken in France at the CERGA 0.9m Schmidt telescope, while southern
  ones are copies from the ESO and SERC surveys.

\begin{table*}[htbp]
\caption{Schmidt fields}
\label{tab:fields}
\begin{tabular}{lrlrlrcc}
     \hline
   Field & \multicolumn{2}{l}{$\alpha$ (2000.0)} &
    \multicolumn{2}{l}{$\delta$ (2000.0)} & Area* &
    \multicolumn{2}{c}{Plates} \\
     \cline{2-5}\cline{7-8}
   & h & m & \degr & \arcmin & sq. degrees & ``blue'' & ``red'' \\
     \hline
     \hline
   3\fh-45\degr   & 03 & 46 & -44 & 40 & 21.57 & SERC 249J & ESO 249F \\
   3\fh-50\degr   & 03 & 32 & -49 & 40 & 21.89 & SERC 200J & ESO 200F \\
   4\fh-45\degr   & 04 & 13 & -44 & 40 & 20.74 & SERC 250J & ESO 250F \\
   4\fh-50\degr   & 04 & 01 & -49 & 40 & 20.38 & SERC 201J & ESO 201F \\
   9\fh+78\degr   & 09 & 22 & +78 & 00 & 20.38 & CERGA 2970J & CERGA 3048F \\
   15\fh+50\degr  & 15 & 36 & +50 & 00 & 20.22 & CERGA 3074J & CERGA 3063F \\
   16\fh+42\degr  & 16 & 31 & +42 & 17 & 20.29 & CERGA 3078J & CERGA 3084F \\
     \hline
\end{tabular}\\
{\small * used in the final catalog.}
\end{table*}

\section{Data processing}
\label{par:dataproc}
\subsection{Digitization}
  The 14 plates were scanned in density mode, with the MAMA microdensitometer, using a 10\mic\ step
  (0.65{\arcsec}).
  The MAMA is able to scan a full plate in a few hours, and provides images with a usable dynamic
  range quite large compared to similar machines like COSMOS or APM: more than 3 in density (a description
  of the microdensitometer and its performances can be found in Berger \etal \cite{berger:al}).
  Because of limitations in acquisition and storage capabilities with the MAMA at that time,
  the image data were acquired as 12{\arcmin}$\times$12{\arcmin} frames, and recombined as
  18{\arcmin}$\times${18\arcmin} images, with an overlap of 6{\arcmin} (which therefore defines the
  maximum size allowed for objects to be reliably measured).

  To insure a maximum reliability of the catalog, only the central ``clean''
  4.5\degr$\times$4.5\degr
  of each field were kept for analysis (ESO and CERGA plates cover only slightly more than 5{\degr}$\times$5{\degr}
  in total, and full coverage was not needed here). The exact areas retained for the final catalogs are
  reported in Table~\ref{tab:fields}.

\subsection{Astrometry}
  The standard MAMA procedure was adopted to calibrate the plates astrometrically (see Berger \etal \cite{berger:al}).
  Between 150 and 300 suitable astrometric standards (PPM catalog, R\"oser \& Bastian \cite{roser:bastian}) can be found
  per plate. A third order fit on projected
  coordinates leads to a mean residual of $\approx0.2$\arcsec, similar to the value obtained when
  comparing bright object positions between the two passbands. However, this figure degrades noticeably
  ($\approx$1\arcsec)
  at the extreme borders of some plates. We had to adjust ESO plates 200 and 249 to their blue
  counterparts SERC~200 and 249 using a neural network mapping\footnote{some other examples of using
  neural networks as an interpolation tool in multidimensional analysis can be found in Serra-Ricart
  et~al. \cite{serra:al} or Bertin \cite{bertinb}.}
  on bright stars to secure blue/red object matching.

\subsection{Detection}
  A dedicated version
  of the SExtractor software (Bertin \& Arnouts \cite{bertin:arnouts}, hereafter BA96) was used in {\tt PHOTO} mode for
  source extraction. The standard 2 pixels (1.3\arcsec) FWHM Sextractor's convolution mask was
  applied in order to improve the
  detectability of low surface brightnesses\footnote{In SExtractor, convolution is only applied
  to the template frame on which objects are detected, not to the image itself; subsequent measurements
  are unaffected by this operation.}. Although emulsion noise becomes strongly correlated on small
  scales (especially on copies), this technique enables one to lower the detection threshold to a secure $1.5\sigma$
  of the sky background fluctuations, corresponding to mean surface brightnesses
  $\mu_{\small \Bj} = 24.9 {\rm \ mag.arcsec}^{-2}$
  (CERGA blue plates), $\mu_{\small \Rf} = 22.8 {\rm \ mag.arcsec}^{-2}$ (CERGA red plates),
  $\mu_{\small \Bj} = 25.9 {\rm \ mag.arcsec}^{-2}$ (SERC blue plates) and
  $\mu_{\small \Rf} = 23.8 {\rm \ mag.arcsec}^{-2}$ (ESO red plates).

  The deblending capabilities of SExtractor give no limitation to the size of objects which can be
  extracted; spiral arms and other peripheral substructures of bright galaxies are not ``separated''
  from the central region. Therefore the catalog is expected to be complete for all objects smaller
  than the overlap between subframes: 6{\arcmin} (corresponding here to $\Bj\approx 12$).

\subsection{Star/galaxy separation}
\label{chap:stargal}
  Star-galaxy separation was performed with a dedicated neural network (Bertin \cite{bertina}) trained on
  a sample made of bright objects classified by eye, and fainter ones from the CCD
  photometric fields (\S \ref{chap:ccd}). The latter were classified automatically using SExtractor's
  tunable neural network classifier (see BA96).
  Briefly, each detected object is translated into a pattern vector containing basic information about its profile
  (7 isophotal areas plus the peak density). This combination of parameters, added to the ability of
  neural networks to deal with complex distributions in multidimensional space, leads to a very robust
  star/galaxy classifier. In particular, objects affected by optical distortions (in the corner of the plates)
  or merging (we chose to classify undeblended pairs as ``stars'' if and only if the brightest component
  is a star) are properly handled. Unfortunately, as we will see, such a set of parameters is not well fitted
  to the classification of very bright stars and galaxies, which show similar profiles.

  A minimum of 500 sample patterns is necessary for the neural
  network to achieve performances close to its asymptotic values (Bertin \cite{bertinb}); we used $\approx 800$
  per field.

  Extensive visual checks and colour histograms show that
  stellar contamination among catalogued galaxies due to misclassifications 
  is less than 4\% for $14<\Bj<19.5$, rising to 10\% at $\Bj \approx 20$ (CERGA plates)
  and $\Bj \approx 21$ (UKST plates). $\Bj = 20$ and $\Bj = 21$ therefore define the upper magnitude
  limits for the northern and southern sets of plates, respectively.
  The estimated loss of galaxies due to misclassifications is similar to
  stellar contamination at faint magnitudes. We can rely on statistics from the identifications of optically bright IRAS
  galaxies (Bertin et~al. \cite{bertin:al}), which contain less than 10\% of objects classified as ``stars'' for $\Bj>16$.
  As a high fraction of IRAS galaxies turn out to be rather compact
  and difficult to separate from point-sources on photographic images
  (Sutherland et~al. \cite{sutherland:al}), this gives us an upper limit of the loss we might expect
  at these magnitudes.
  It is more difficult to estimate it precisely at the bright end of our counts (($\Bj<16$),
  because stars outnumber galaxies by a factor $> 50$. We examined 600 detections with $\Bj\approx 15.5$,
  classified as stars, and found indeed that 4 of them were galaxies. There seems therefore to be some loss of galaxies
  at the bright end of our catalog, and counts below $\Bj=16$ (or $\Rf=14.5$) should only be considered as lower limits.

\subsection{Photometry}
  \subsubsection{Estimation of magnitudes}
  The way SExtractor estimates ``total'' magnitudes on CCD images is described in details in BA96.
  In the ``{\tt PHOTO}'' mode, the program applies a density-to-intensity
  transformation (see \S \ref{chap:calib}) before summing pixel fluxes.
  Briefly, for each object, two kinds of magnitudes are computed. The first one is an improvement
  of Kron's (\cite{kron}) method: it measures the
  flux integrated in an elliptical aperture whose size and shape are function of the object's profile.
  The second one is isophotal, and corrects for the fraction of flux lost in the wings
  (assuming a gaussian profile) as in the APM survey (Maddox et al. \cite{maddox:alc}). As this second
  method is more subject to biases than the first one, we take the aperture magnitude as
  an estimate of ``total'' magnitude, unless it is suspected to be contaminated by the presence
  of neighbours by more than 0.1 mag; in which case we rely on corrected isophotal magnitude.
  Such a situation occurs in less than 20\% of the cases in our images; SExtractor's total magnitude
  is thus essentially an aperture magnitude.

\begin{figure}[htbp]
  \centerline{\psfig{figure=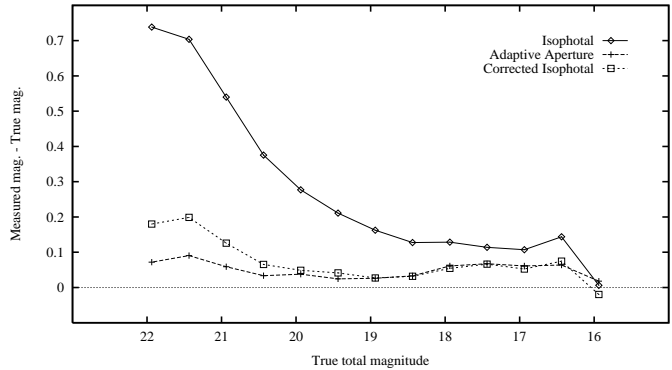,width=8.8cm,angle=-90}}
  \caption[]{
            Prediction of the mean flux lost (expressed in magnitude)
            by SExtractor's isophotal, corrected isophotal and adaptive aperture magnitudes,
            as a function of the ``true'' total magnitude. These estimates are from simulated
            Schmidt plates images, generated as described in BA96.
            The simulations are done in the blue photographic passband; they assume a logarithmic
            response of the emulsion and image parameters typical of CERGA Schmidt plates, with a seeing
            FWHM of 2{\arcsec}.
            The apparent drop seen at $\Bj = 16$ is an artefact due to poor statistics.
            }
      \label{fig:sexmag}
   \end{figure}

  The behaviour of SExtractor's magnitudes on simulated Schmidt plate images is shown in Fig. \ref{fig:sexmag}.
  The predicted fraction
  of flux measured for galaxies is remarkably constant with magnitude, even beyond the completeness limit.
  As can be seen, the mean offset one needs to apply to SExtractor's magnitude to get an
  estimate of ``total magnitude''  is $\approx -0.06$~mag.

\subsubsection{Linearization of the photographic response}
\label{chap:calib}
  In {\tt PHOTO} mode, SExtractor transforms densities $D$ to intensities $I$ during image analysis
  assuming a logarithmic response of the emulsion in the ``linear part'':
\begin{equation}
\label{eq:calib}
I = I_0.10^{\frac{D}{\gamma}}
\end{equation}
  where $\gamma$ defines the contrast of the emulsion (as measured with MAMA), and
  $I_0$ is the intensity zero-point. Both parameters were adjusted for each plate by minimizing the
  $\chi^2$ on ``total magnitudes'' estimated by SExtractor for a set of CCD standard galaxies.
  For all plates but one we found $2<\gamma<3$. Differential desensitization
  and telescope vignetting were compensated for by substracting the local background density from $D$
  before applying the calibration law (Eq. \ref{eq:calib}). This correction is perfectly valid here as no
  interstellar emission is
  expected in these fields, and reduces large-scale sensitivity variations to $\la0.1$~mag
  (Bertin \cite{bertinb}).

  Finally, for galaxies, a linear least-square fit
  was applied separately to the magnitude scale in each field to correct for residual differences
  of $\approx\pm3$\% in the slope of the relation between photographic and CCD magnitudes.
  Bright stellar images are affected by a mix of photographic saturation, complex chemical proximity
  effects and, above all, light diffusion inside the emulsion during the scanning process.
  Stars thus need a specific calibration which was carried out
  using polynomial and neural network fits between their photographic and CCD magnitudes.

  Eq. (\ref{eq:calib}) provides a good fit to
  photographic calibration curves of sky-limited exposures measured with MAMA
  (Moreau \cite{moreau}, Bertin \cite{bertinb}), although it does not take into account
  possible saturation effects. Thanks to the fairly large dynamic range of MAMA,
  saturation affects our galaxy magnitudes to a lesser extent than e.g. APM or COSMOS scans.
  Nevertheless, southern {\em copy} plate images unambiguously show some
  signs of saturation on cores of galaxies with $\Bj\la17$. We therefore expect that both
  systematic and random errors increase significantly towards brighter magnitudes on these plates
  (see Metcalfe \etal \cite{metcalfe:alc} for a discussion about the consequences of saturation
  on galaxy photometry).

\subsubsection{CCD calibration and colour equations}
\label{chap:ccd}
\begin{figure*}[htbp]
  \centerline{\hbox{\psfig{figure=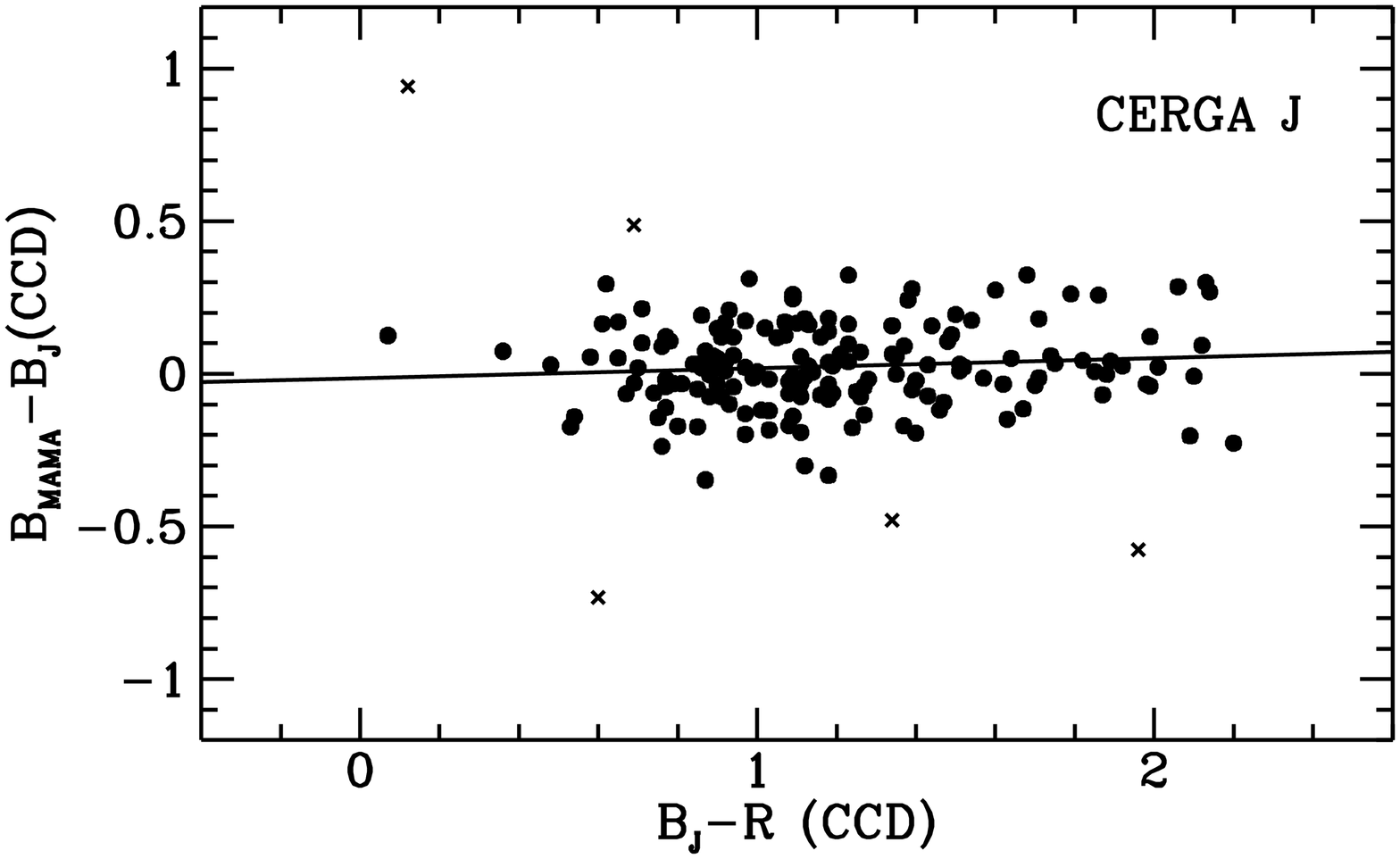,width=9cm,angle=-90}
                    \psfig{figure=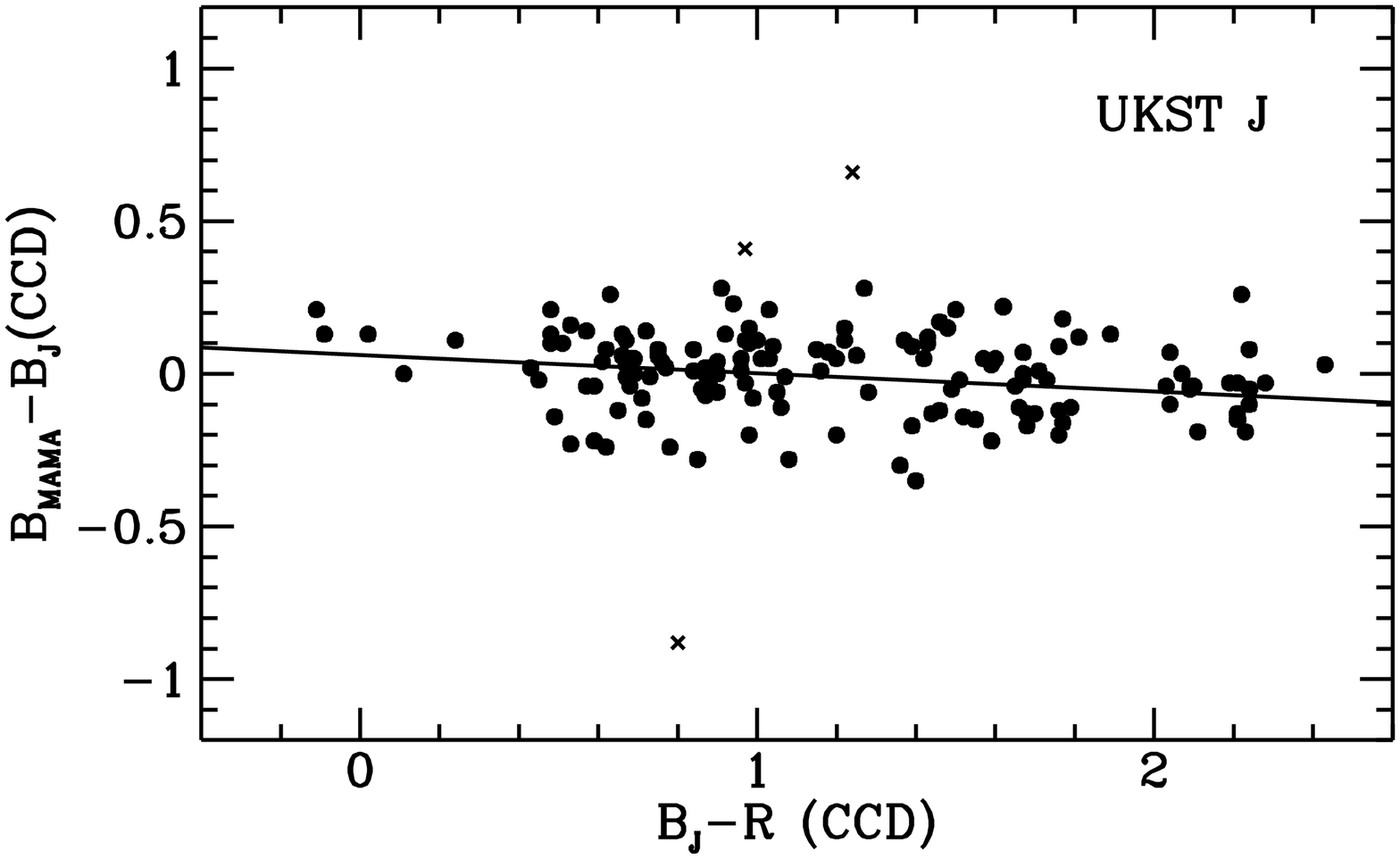,width=9cm,angle=-90}}}
  \centerline{\hbox{\psfig{figure=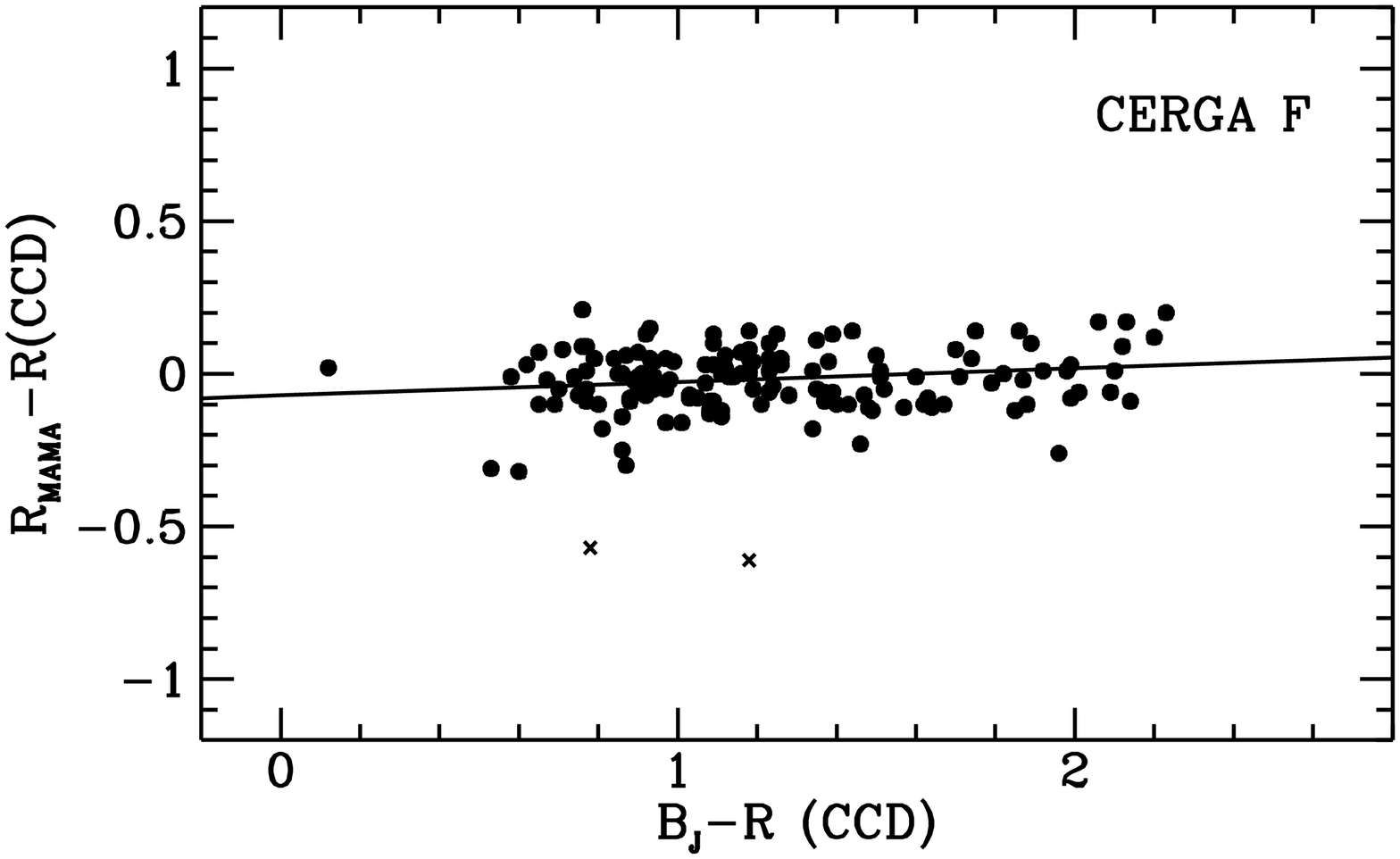,width=9cm,angle=-90}
                    \psfig{figure=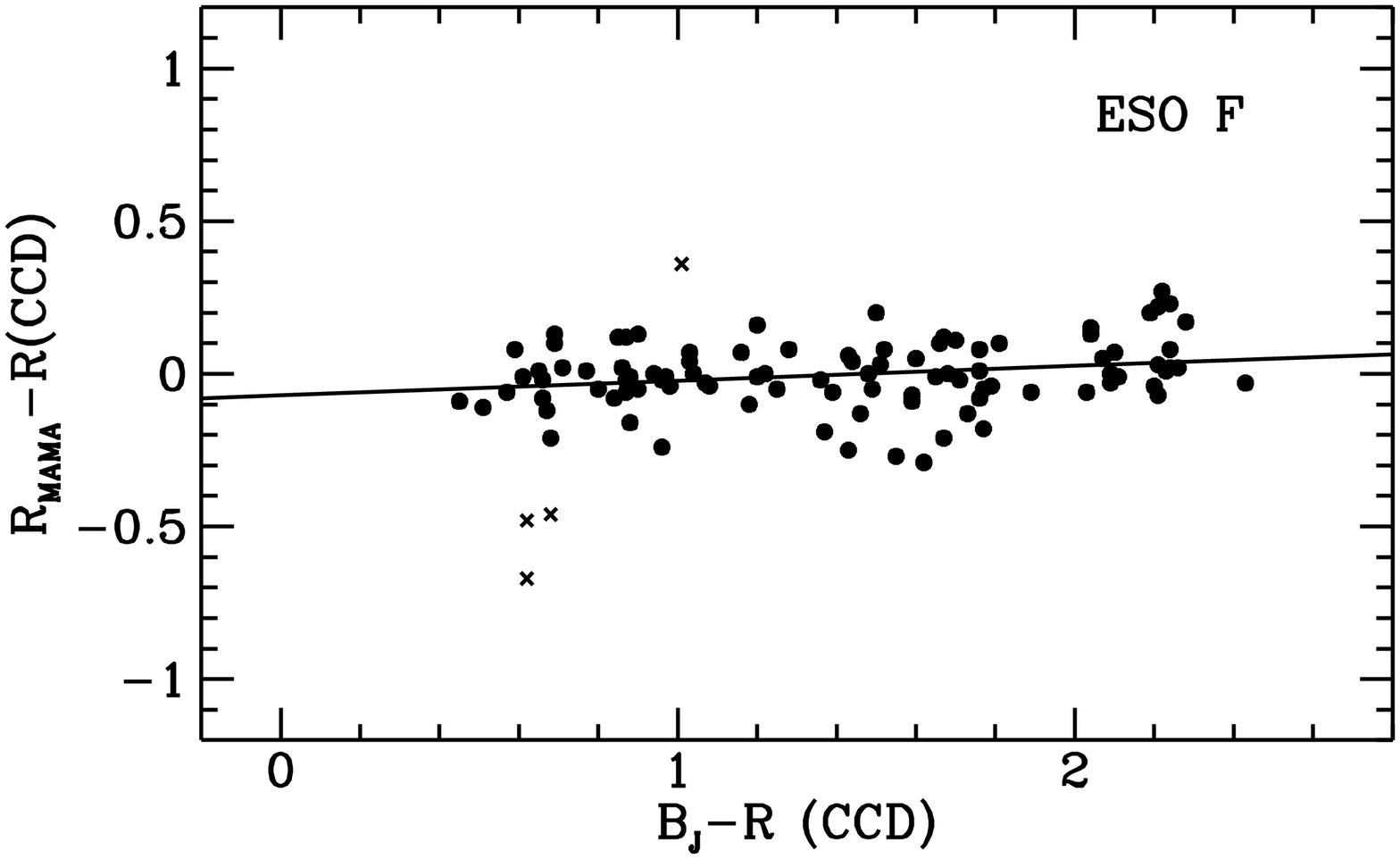,width=9cm,angle=-90}}}
  \caption[]{
            Differences between ``pure'' (but calibrated) photographic and CCD magnitudes as a function
            of the $\Bj-\R$ colour index for the brightest standards: $\Bj<20$ (CERGA blue), $\Bj<21$
            (UKST blue), $\R<18$ (CERGA red) and $\R<19$ (ESO red). The standards contain about 2/3 of stars
            and 1/3 of galaxies (which show similar behaviour).
            A few points (crosses) deviate by more than $3\sigma$ from the mean
            relation and have not been taken into account for the chi-square fit (solid line).
            }
      \label{fig:colourcomp}
   \end{figure*}

  A total of 30 photometric CCD fields ($\approx$~4 per Schmidt field,
  centered on groups of bright galaxies) were taken essentially at the OHP 1.2m and the ESO
  2.2 and 3.6 meter telescopes, and matched to Cousins B and R magnitudes. SExtractor was also used
  at all stages of the CCD data reduction, leading to ``total magnitudes'' for about 1000 objects.
  As northern and southern plate sets have been taken with slightly different filter combinations,
  we naturally expect differences in their related colour equations. However we need to keep the survey
  as close as possible to a uniform (and as standard as possible) passband system.
  Following Blair \& Gilmore (\cite{blair:gilmore}), we define here the blue passband $\Bj$ as
\begin{equation}
\label{eq:bjcolour}
\Bj \equiv \B -0.28(\B-\V) \ \ \approx \B-0.19(\B-\R)
\end{equation}
which is the transformation also adopted in the APM survey.
The $\Rf$ band is defined as
\begin{equation}
\Rf \equiv \R
\end{equation}
  $\B$, $\V$ and $\R$ are in the Cousins system \footnote{Although our $\Rf$ is photometrically equivalent
  to Cousins $\R$ within the measurement errors, throughout this paper we denote red
  {\em photographic} magnitudes ``$\Rf$'' instead of ``$\R$''.}. Figure \ref{fig:colourcomp} shows
  the distribution of the residuals as a function of the $\Bj-\R$ CCD colour index for each
  of the 4 types of plates. Although they are moderate, one can see systematic colour
  effects. The ``best fit'' colour coefficients are reported in Table \ref{tab:coloureqs}.
  For CERGA blue plates the results are in good agreement with the estimates of Majewski (\cite{majewski}),
  who finds $\B_{\mbox{\tiny IIIaJ+GG385}} = \B - 0.23 (\B-\V)$. We find a stronger
  colour coefficient for the SERC passband than the one traditionally used at the APM
  ($\B_{\mbox{\tiny IIIaJ+GG395}} = \B - 0.28 (\B-\V)$), and conflicting even more
  with the COSMOS ($\B_{\mbox{\tiny IIIaJ+GG395}} = \B - 0.23 (\B-\V)$)\footnote{It seems that through
  the years some confusion has often been made between the IIIaJ+GG385 and IIIaJ+GG395 combinations (see for instance
  Shanks \etal \cite{shanks:al}), and yet these prove to lead to somewhat different (although close) passband definitions.}.
  However our estimate is in remarkable agreement with the recent and accurate measurements
  by Metcalfe et al. (\cite{metcalfe:alc}), which give
  $\B_{\mbox{\tiny IIIaJ+GG395}} = \B_{\mbox{\tiny IIaO}} - 0.35 (\B-\V)$
  and $\B_{\mbox{\tiny IIaO}}\approx B$.

  Lastly, we find in average the red photographic passbands slightly {\em bluer} than the Cousins $\R$.
  This contradicts the traditional correction established
  --- photoelectrically, and then extrapolated to photography --- by Couch \& Newell (\cite{couch:newell}),
  who proposed a response for the combination IIIaF+RG630 {\em redder} than Cousins $\R$.
  Note that more recent determinations (Cunow \& Wargau
  \cite{cunow:wargau}) also support the idea that the Couch \& Newell equation is inappropriate.

\begin{table*}[htbp]
\label{tab:coloureqs}
\caption{Photographic passbands and their link to standard photometric systems}
\begin{tabular}{lccc}
     \hline
   Plate type & Emulsion+filter & \multicolumn{2}{c}{Colour equations} \\
     \hline
     \hline
   CERGA ``blue''   & IIIaJ + GG385 & $\Bj + (0.033 \pm 0.015)\,(\Bj - \R)$ & $\B-0.24(\B-\V)$ \\
   UKST ``blue''   & IIIaJ + GG395 & $\Bj-(0.060 \pm 0.023)\,(\Bj - \R)$ & $\B-0.35(\B-\V)$ \\
   CERGA ``red''   & IIIaF + RG610 & $\R+(0.044 \pm 0.017)\,(\Bj - \R)$ & $\R+0.036(\B-\R)$ \\
   ESO ``red'' & IIIaF + RG630 & $\R+(0.048 \pm 0.019)\,(\Bj - \R)$ & $\R+0.039(\B-\R)$ \\
     \hline
\end{tabular}
\end{table*}

  In the end, the photographic magnitudes from each plate where corrected using the equations
  of Table \ref{tab:coloureqs} to yield magnitudes in a unified system.
  Given the smallness of the colour coefficients, the resulting degradation in photometric accuracy is
  negligible.
  The {\it rms} residual
  of the calibration above the completeness limit ranges between 0.07 and 0.17 mag (including the
  contribution from large scale inhomogeneities). Assuming that the intensity scale is perfectly
  linearized, the formal uncertainty on the individual plate zero-points would be $\approx 0.02$~mag.
  But such an assumption is by far too optimistic with photographic plates.
  Given the number of bright standard galaxies per plate (between 1 and 3 per magnitude), and the {\rms}
  uncertainty on their magnitudes (about 0.15 mag), we estimate the individual zero-point systematic
  errors to be $\la 0.1$ mag in the range $15<\Bj<21$, and $14<\Rf<19.5$. As Figure \ref{fig:magdif}
  shows, below these fluxes the photometric errors grow rapidly.

\subsubsection{Comparison with other photometries}
\begin{figure*}[htbp]
  \centerline{{\hbox{\psfig{figure=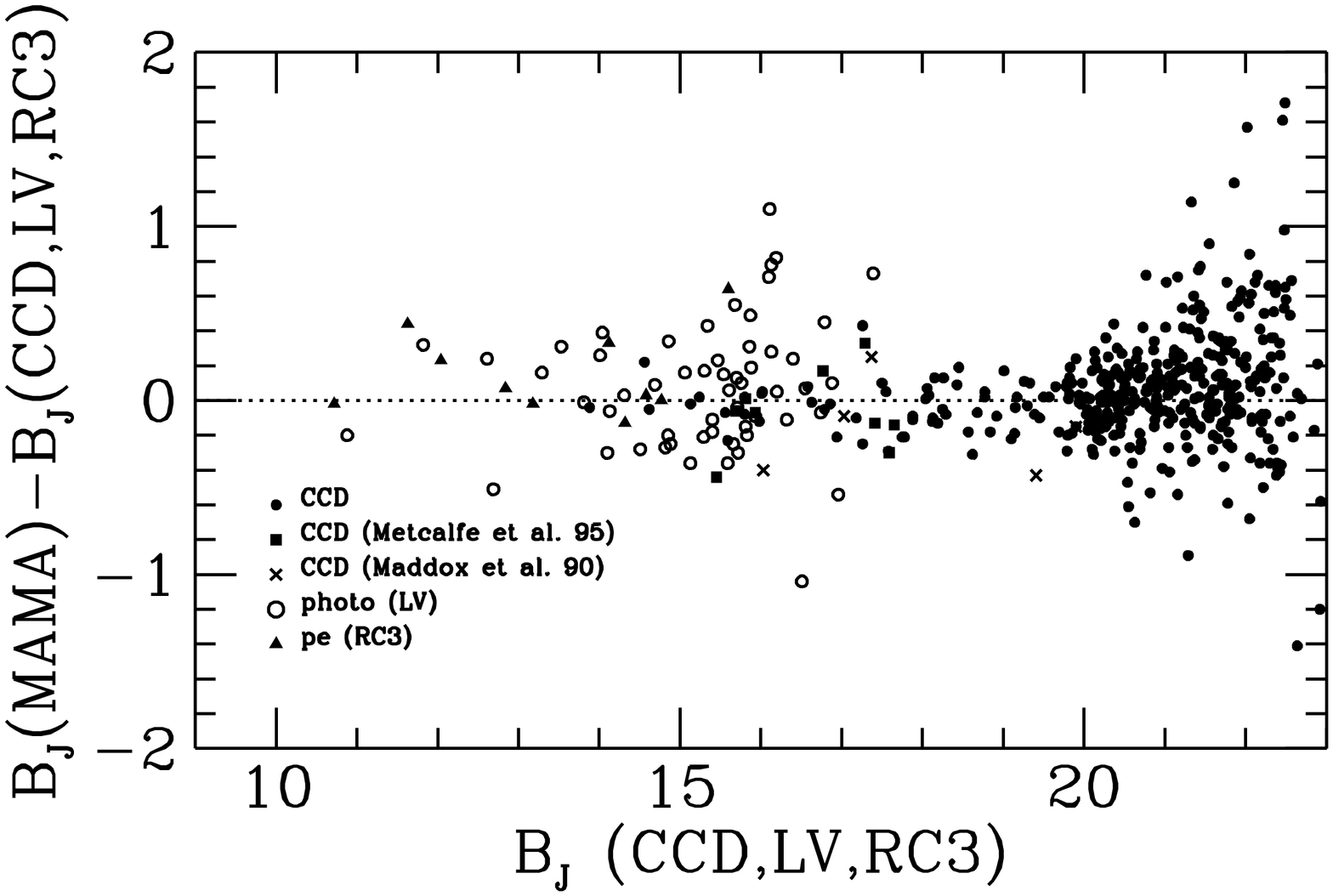,width=8.8cm,angle=-90}}
		\psfig{figure=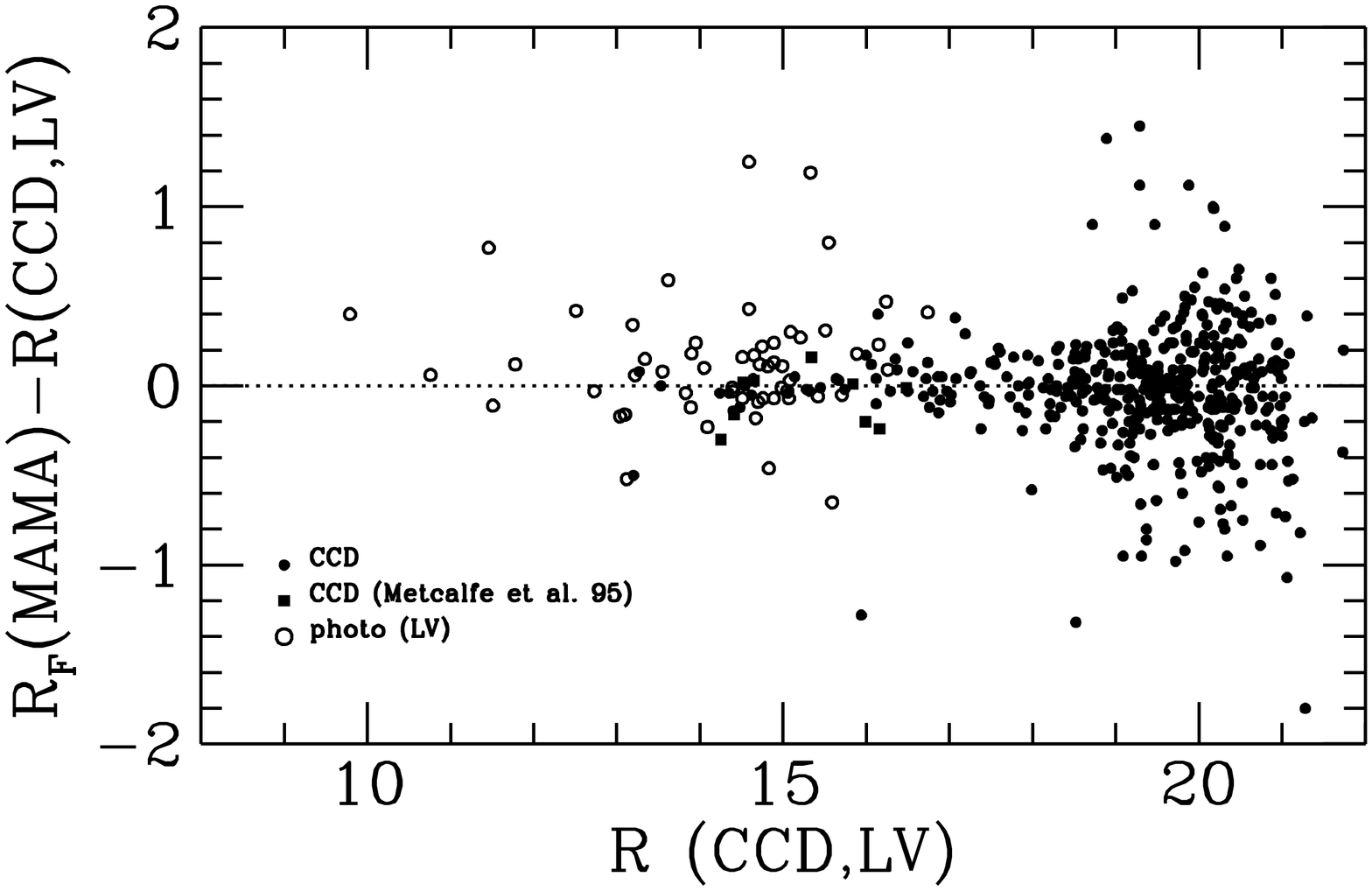,width=8.8cm,angle=-90}}}
  \caption[]{
            Difference between photographic and CCD magnitude for standard galaxies as a function
            of CCD magnitude. Photographically unsaturated stars (with $\Bj>20$ or $\Rf>18$) are added at the
            faint end. Also displayed are bright galaxies measured by different authors.}
      \label{fig:magdif}
   \end{figure*}

  Figure \ref{fig:magdif} shows the differences between photo
  and CCD magnitudes for our CCD standard galaxies, as well as the comparison with the photometry
  of bright galaxies by several authors.
  Nine of the galaxies measured by Metcalfe et~al. (\cite{metcalfe:alc})
  are found in our catalog; we find mean differences (MAMA-Metcalfe) $\Delta\Bj = -0.04 \pm 0.08$,
  and $\Delta\Rf = -0.07 \pm 0.05$. Note that 3 of these galaxies also figure among our CCD standards;
  in both $\B$ and $\R$, our magnitudes are in agreement with theirs within 0.02 mag.

  Although their photographic photometry does not reach the same accuracy and is more subject to biases
  than CCD or photoelectric standards,
  galaxies from the LV catalog (Lauberts \& Valentijn \cite{lauberts:valentijn}) have proven to be in
  good agreement with the RC3 system (Paturel et~al. \cite{paturel:ala}), and can be used to trace
  an eventual large systematic trend at bright magnitudes ($\Bj<16$).
  Over the 4 southern fields we find mean differences (MAMA-LV) $\Delta\Bj = +0.10 \pm 0.05$ and
  $\Delta\Rf = +0.06 \pm 0.05$. These discrepancies might be interpreted as some loss of flux
  at the bright end ($\Bj\approx 15$) of our catalog. However, in the same magnitude range,
  ``CCD galaxies'' photometered here or by Metcalfe et~al. do not show this offset, which could argue
  for a small, local zero-point error in the LV magnitudes.

  Some bright galaxies ($\Bj\la 15$) from the northern plates do also have photoelectric $\B$ and $\V$
  ``total'' magnitudes in the RC3 catalog. From these we get a mean difference (MAMA-RC3)
  $\Delta\Bj = +0.15 \pm 0.08$ which reveals the influence of plate saturation at the bright end of
  our catalog.

  Finally, 5 of the galaxies photometered with a CCD by Maddox et~al. (\cite{maddox:alc}) to calibrate the
  APM survey lie in our field 201 and give an offset (MAMA-APM) $\Delta\Bj = -0.16 \pm 0.14$.  More than
  this significant offset, the unexpectedly large dispersion of magnitudes --- almost 0.3, that is,
  as much as with the photographic LV sample! --- 
  casts some doubt over the reliability of the Maddox et~al. calibration set for this plate.

  In conclusion, we believe our magnitude scale to be free from any large systematic error,
  except brightwards of $\Bj\approx15$, where fluxes might possibly be underestimated by $\approx0.1$~mag.

\subsection{Merging of catalogs}
  The ``blue'' and ``red'' catalogs were cross-identified in alpha, delta, to yield a unique
  two-color catalog.
  Differences in seeing or image quality, as well as imbricated detections (i.e. stars lying
  on disks of galaxies) were handled through a complex matching algorithm taking into account
  the shape of detected objects.
  This procedure leads to a highly reliable catalog,
  virtually suppressing all false detections like pieces of hair, satellite trails, optical ghosts
  or spikes around bright stars (these are generally classified as galaxies). 
  One might fear some loss of objects with extreme colours in the final
  catalog; however the fraction of non-paired detections is almost constant with magnitude,
  and is about 1\% (most of which are spurious), rising to 5\% in the last half-magnitude bin
  imposed by star/galaxy separation.

  \begin{figure*}[htbp]
  \centerline{\hbox{\psfig{figure=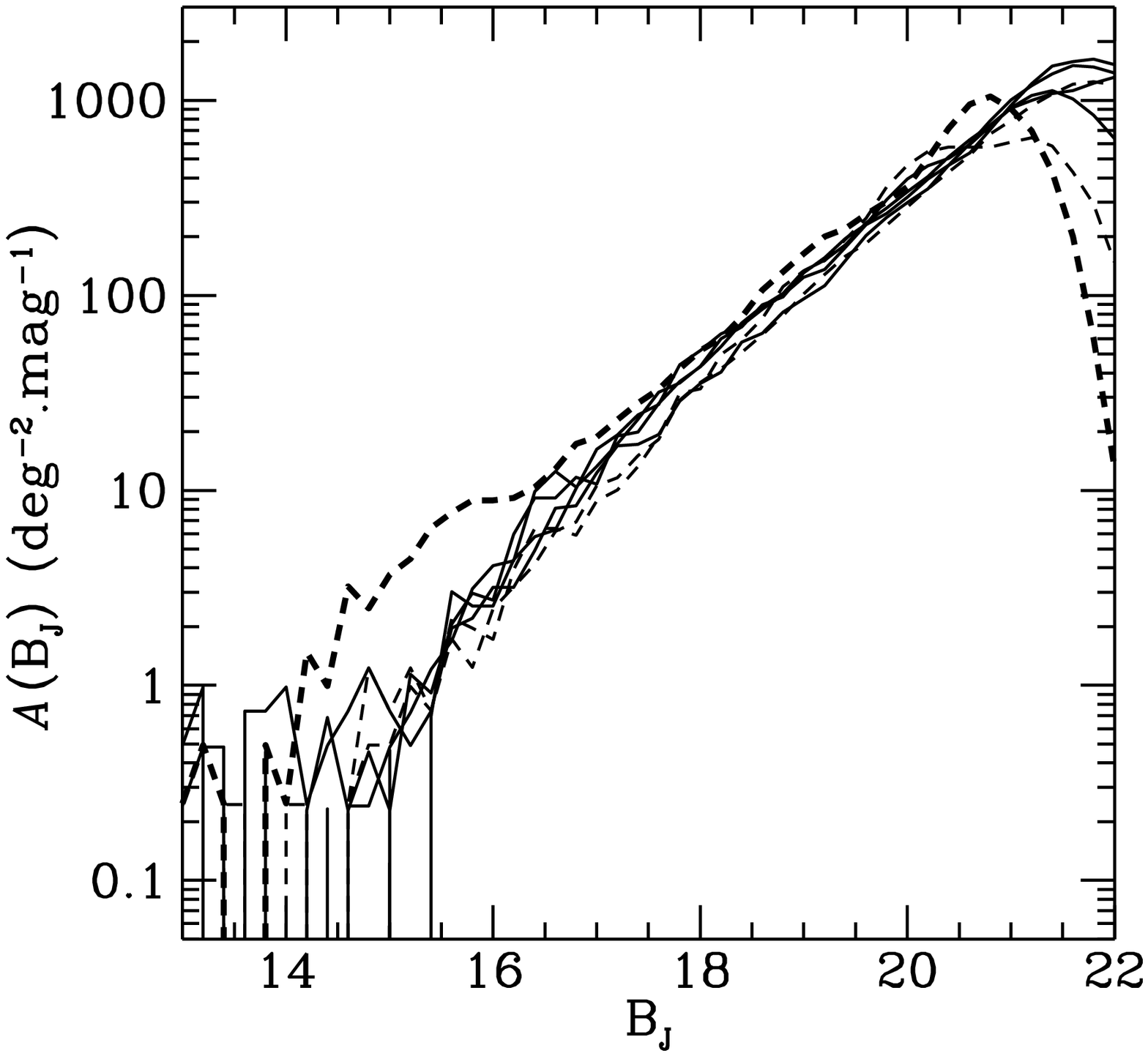,width=8.8cm}
		\psfig{figure=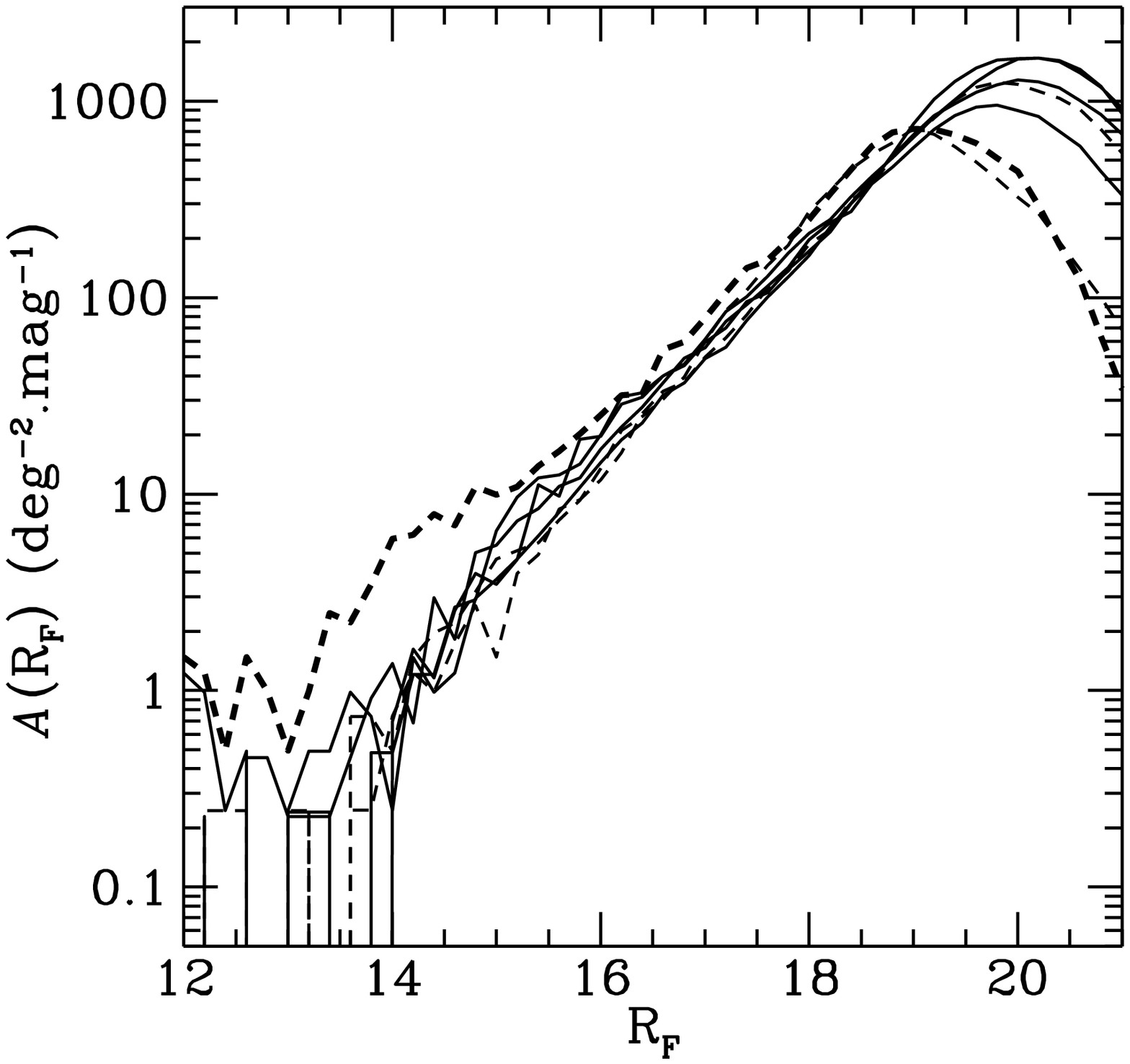,width=8.8cm}}}
  \caption[]{
      ``Raw'' differential number counts in $\Bj$ and $\Rf$ for each of the Schmidt fields (continous line
      = southern fields; dashed lines = northern fields). Note the excess of galaxies in the northern
      field 16\fh+42{\degr} (thicker lines), especially at bright magnitudes.}
   \label{fig:multicounts}
   \end{figure*}

\section{Galaxy number counts}
\label{par:numbercounts}
\subsection{Results}
\begin{figure*}[htbp]
  \centerline{\hbox{\psfig{figure=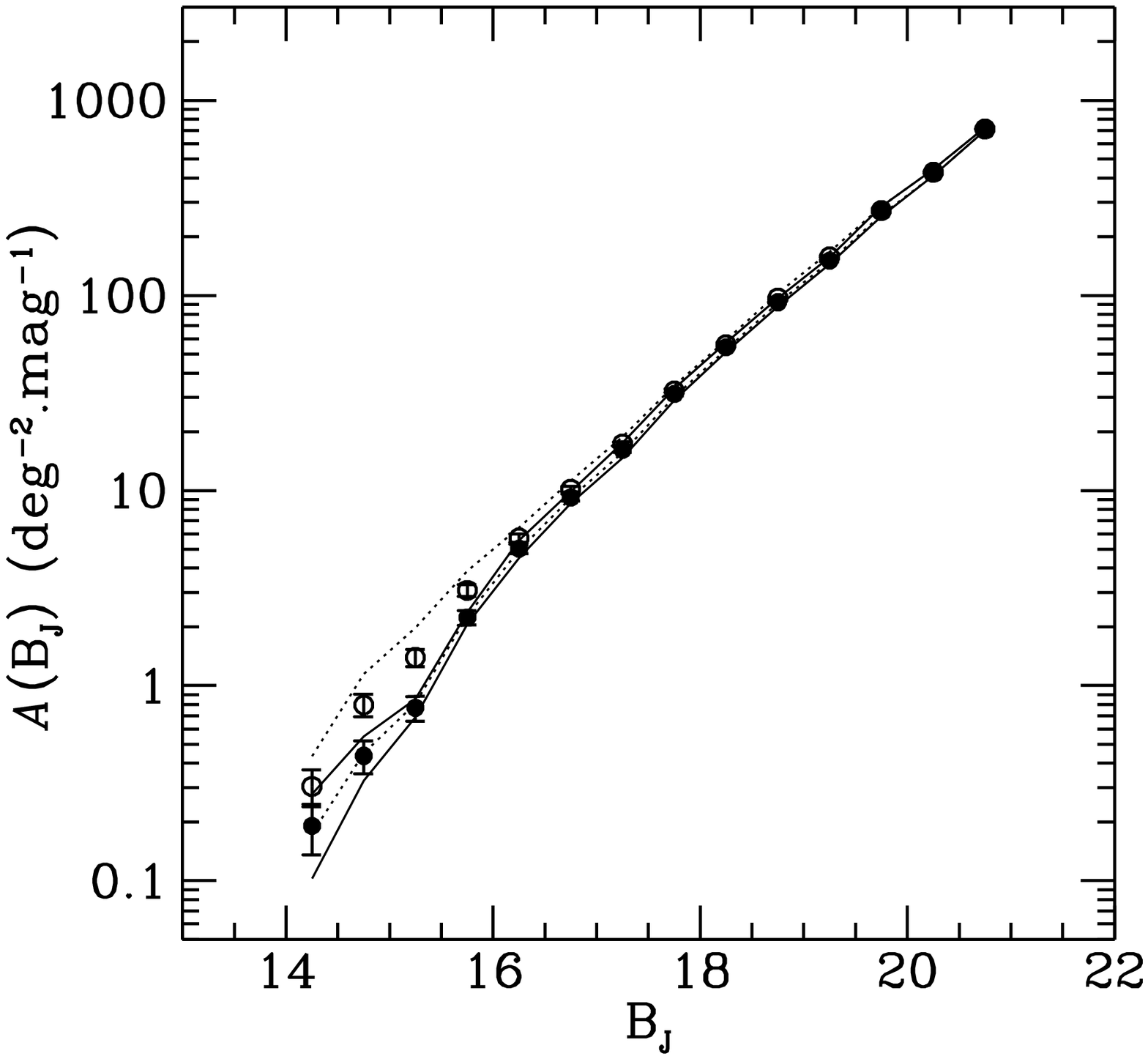,width=8.8cm}
		\psfig{figure=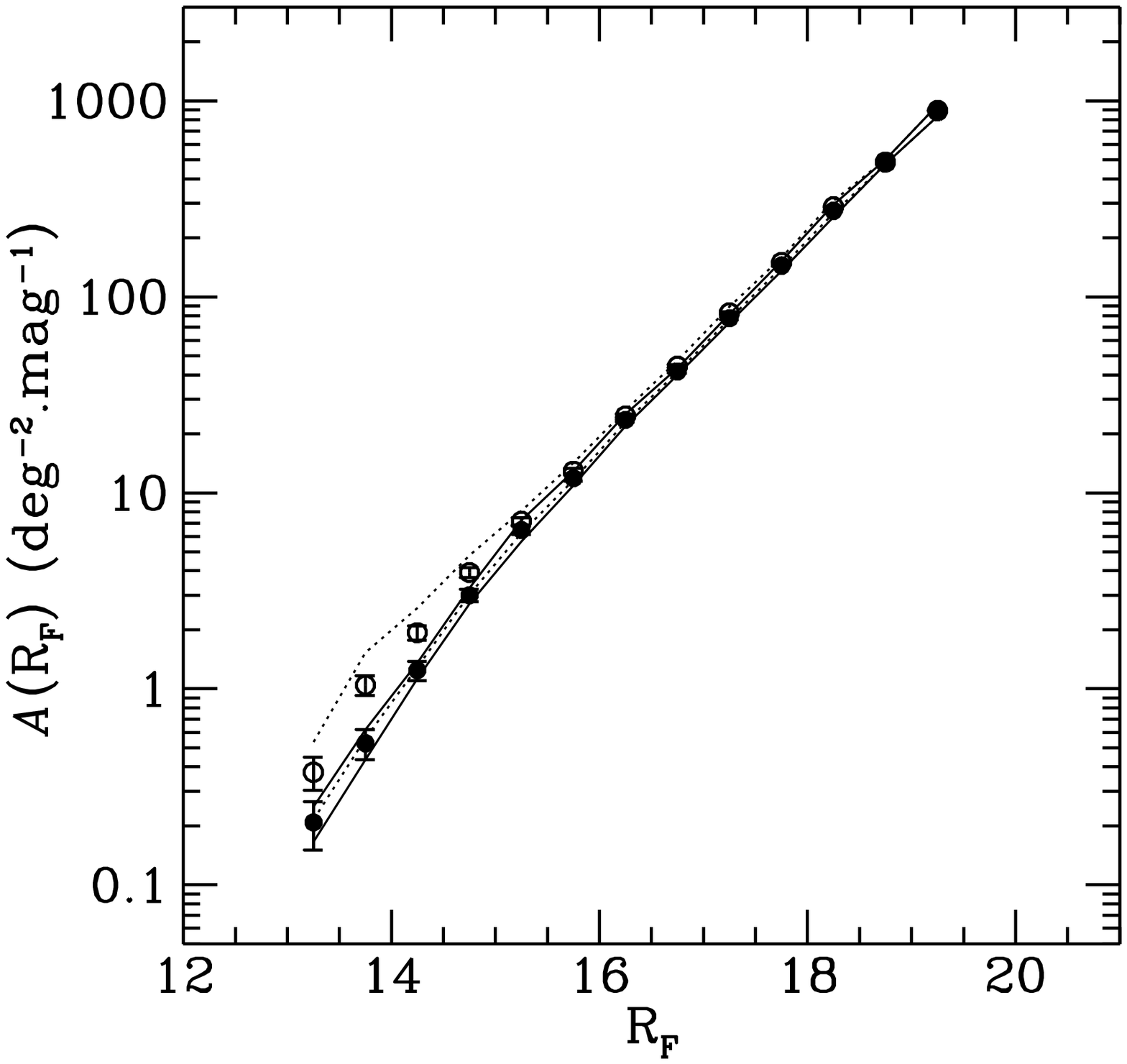,width=8.8cm}}}
  \caption[]{
      Total differential number counts in $\Bj$ and $\Rf$ in our catalog, including (open circles) or
      not (filled circles) the 16\fh+42{\degr} field. Error bars indicate poissonian ($1\sigma$) uncertainties whereas
      lines bracket the \rms\ excursion of the total counts assuming the galaxy density is uncorrelated
      from plate to plate.}
   \label{fig:totcounts}
   \end{figure*}

The differential number counts are shown for each field in Fig.\ref{fig:multicounts}. One can
notice a great homogeneity over the whole magnitude range between the plates, except for the
16\fh+42{\degr} field. In fact this field contains no less than 12 Abell clusters, and is
part of the Hercules supercluster. One can estimate the variance in number counts at a given depth on
a Schmidt plate from the Maddox et al. (\cite{maddox:ala}) angular two-point correlation function
(e.g. Peebles 1980). We find that the projected density of galaxies in the 16\fh+42{\degr} field
exceeds the mean one from the other fields by a factor of $\ga 4$ times the expected deviation
at $\Bj=15$, reducing to $\approx 2$ for $\Bj>17$. This field is therefore one of the densest in
the sky at the bright end of our catalog, and including it in the total number counts
shown in Fig.\ref{fig:totcounts} severely changes the slope at $\Bj \la 16$. For this reason we decided
to discard it from the samples for the statistical analysis of number counts, remembering however
that we might underestimate in this way the true density of bright galaxies.

\subsection{Comparison with previous counts}
\label{chap:countcomp}
Bright galaxy counts published so far were all conducted on photographic material, in quite similar
passbands. This enables us to compare easily our results with previous work. The transformations we applied
are described below.

Many of the existing bright galaxy counts (Shanks et~al. \cite{shanks:al} , Stevenson et~al. \cite{stevenson:al}, Heydon-Dumbleton
et~al. \cite{heydon:al}) make use of the $b_j$ passband definition established by Kron (\cite{kron}), and
which transforms to our $\Bj$ through
\begin{equation}
\Bj = b_j - 0.05
\end{equation}
assuming $\langle \B-\V \rangle = 1.0$. The recent counts of Weir et~al. (\cite{weir:al}) were done in the Gunn-$g$ passband,
for which they give the approximate transformation
\begin{equation}
\Bj \approx g - 0.5
\end{equation}
to the APM survey magnitude system (Maddox et~al. \cite{maddox:alc}), which is identical to ours.

Gunn-$r$ observations (Sebok \cite{sebok}, Picard \cite{picardb}, Weir at~al. \cite{weir:al}) were transformed
to the Cousins $\R$ system using
\begin{equation}
\R = r + 0.341 
\end{equation}
from a fit to 16 Landolt standards, giving a standard error of only 0.014 mag
(P.Fouqu\'e, private communication). Shanks et~al. (\cite{shanks:al}) $r_F$ magnitudes were converted with
\begin{equation}
\R = r_F - 0.09
\end{equation}
inserting $\langle \B-\R \rangle = 1.5$ in their equation (5).

\begin{figure*}[htbp]
  \centerline{\hbox{\psfig{figure=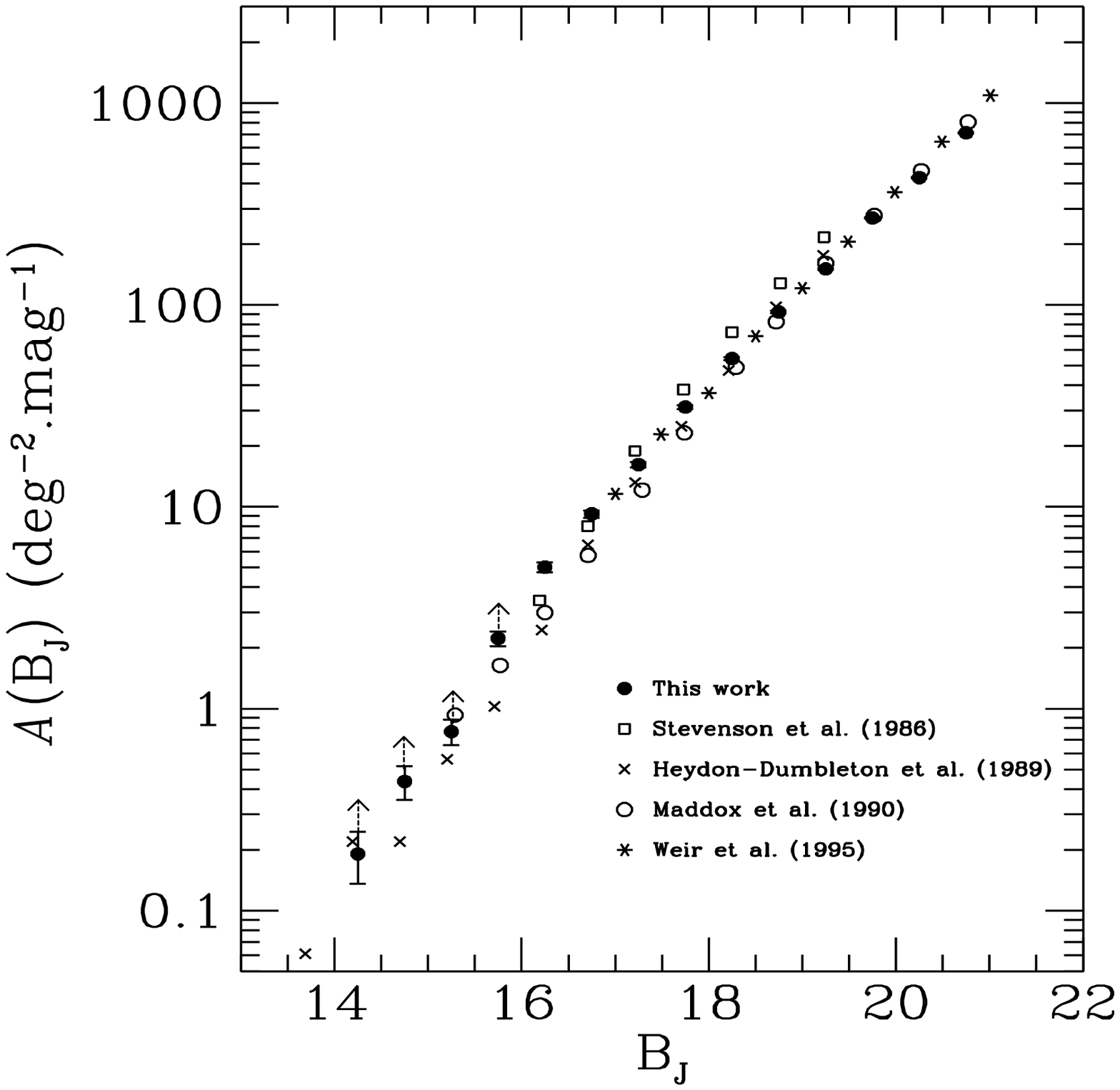,width=8.8cm}
		\psfig{figure=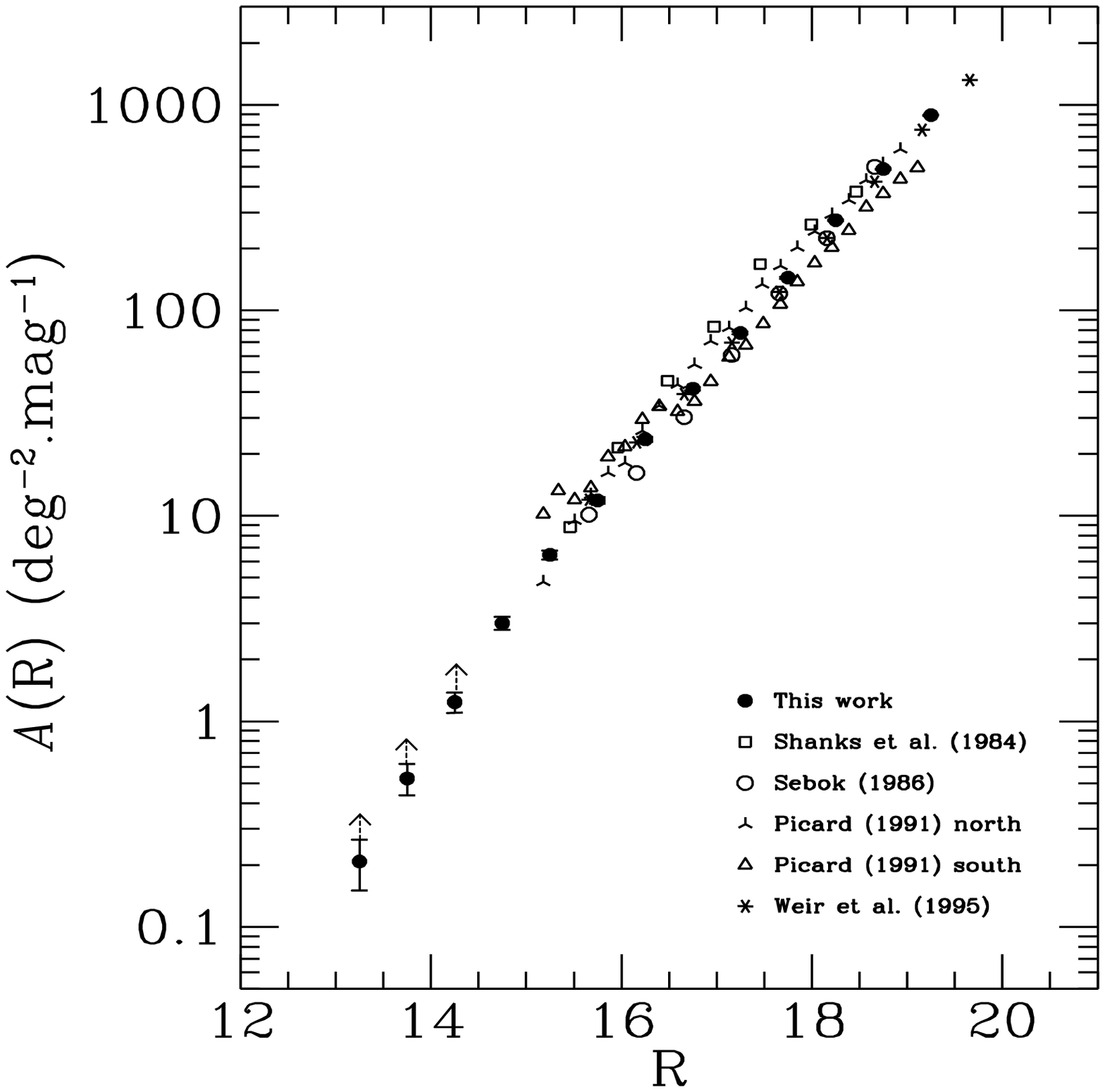,width=8.8cm}}}
  \caption[]{
       Comparison of published bright galaxy counts converted to our $\Bj$,$\R$ photometric system. Arrows
       indicate counts underestimated because of incompleteness (\ref{chap:stargal}).}

   \label{fig:countcompare}
   \end{figure*}

As shown on Fig. \ref{fig:countcompare}, our counts agree well with those from other studies,
especially with the preliminary counts done on POSSII plates (Weir et~al. \cite{weir:al}), despite the
uncertainties in the conversion from their photometric system to ours. In $\R$, the major discrepancies
observed are with Shanks et al. (\cite{shanks:al}), and the Picard (\cite{picardb}) northern and southern counts,
although our results lie in-between. The Schanks et al. counts are based on only one Schmidt plate, so the fact that
they show an offset is perhaps not surprising.
As pointed out by Weir et al. (\cite{weir:al}), it is yet unclear whether the difference between the Picard counts
originates from the presence of very large scale structures, as claimed by Picard (\cite{picarda}),
or are simply caused by unexpected photometric errors in his photometry. We note here that one of our plates
lies at about 20\degr from Picard's northern field, and that no particular enhancement in galaxy density is
seen there.

At their faint end ($20<\Bj<21$), our blue counts are about 10\% lower than those of Weir et~al. (\cite{weir:al}) and Maddox et~al.
(\cite{maddox:ald}); however the raw counts (Fig. \ref{fig:multicounts}) do not indicate any obvious drop in completeness at that level.
As a matter of fact, two points should be considered here: (1) half of the difference (5\%) can be explained by detections
that were not matched between the two colours and dropped in the final catalog (although there is no proof that these are real objects),
and (2) Weir et~al. (\cite{weir:al}) showed on simulations that both their counts and those derived by the APM survey might be
slightly overestimated (by about 10\%) for $\Bj\ga 20$. Therefore it is likely that this difference at the faint end between
our counts and others should not be interpreted as something significant, but only as a consequence of a different data processing.
This is confirmed by the recent medium-deep CCD counts by Arnouts et~al. \cite{arnouts:al}, which are in perfect agreement with ours over
this magnitude range, in both colours.

But the most interesting discrepancy in this comparison is the one seen with some other galaxy counts at bright
magnitudes. At $\Bj=16.5$ for instance, we count nearly twice as many galaxies as do Maddox et~al.
(\cite{maddox:ald}) or Heydon-Dumbleton et~al. (\cite{heydon:al}). The recent Weir et~al. (\cite{weir:al}) counts
stay however in perfect agreement with ours down to their brightest limit.
As the APM survey, because of its statistical weight, is traditionaly used to
normalize galaxy counts models at bright magnitudes, it is worth investigating further what may be
the origin of this difference. Recently, Metcalfe et al. (\cite{metcalfe:alc}) have discovered some scale
error on the $17 \la \Bj \la 18$ domain in the APM galaxy magnitudes. The APM survey shares four Schmidt fields
in common with us, and the corresponding part of the catalog, down to $\Bj = 21$ was kindly provided by J. Loveday
for comparison.

  \begin{figure*}[htbp]
  \centerline{\hbox{\psfig{figure=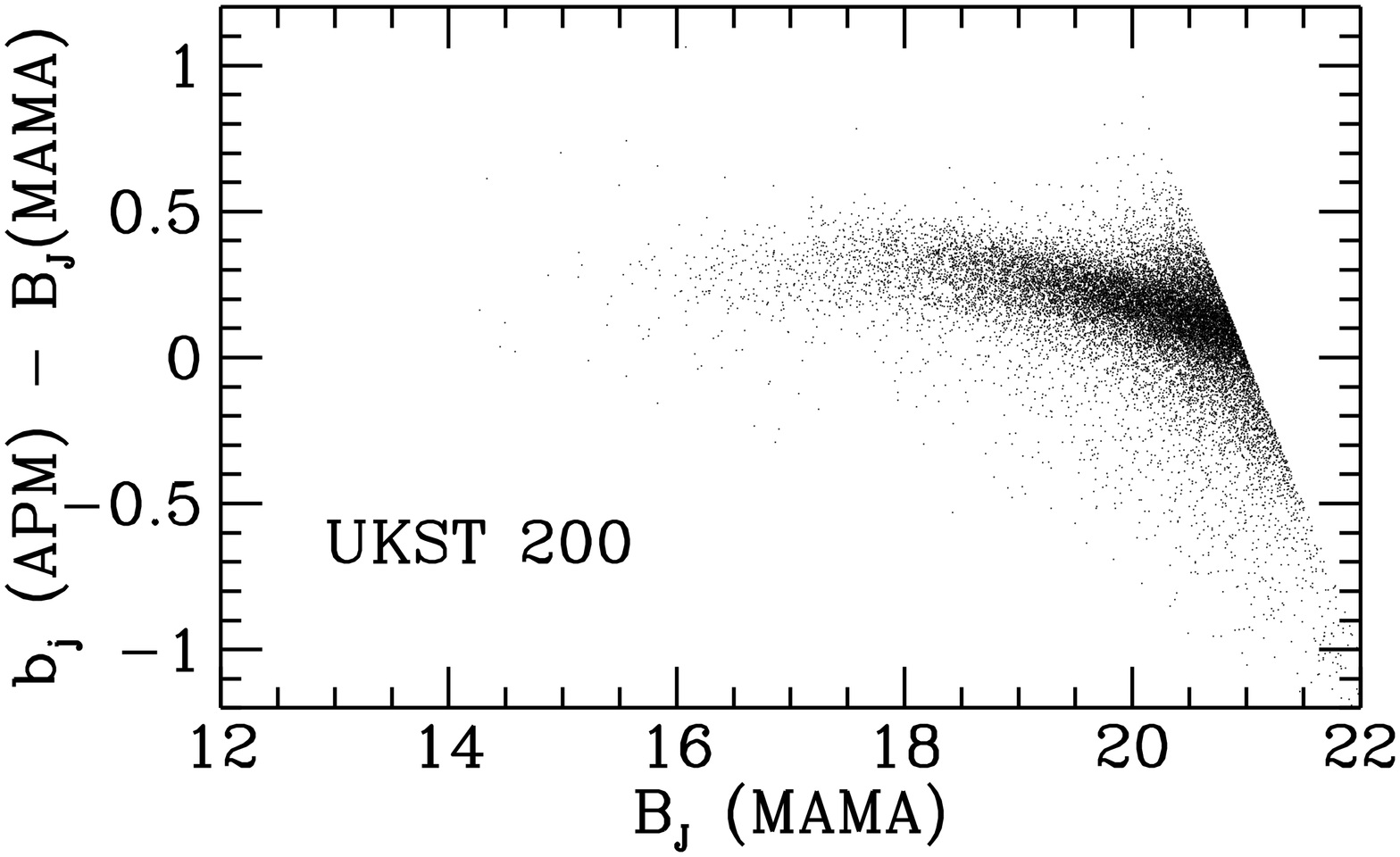,width=8.8cm,angle=-90}
                    \psfig{figure=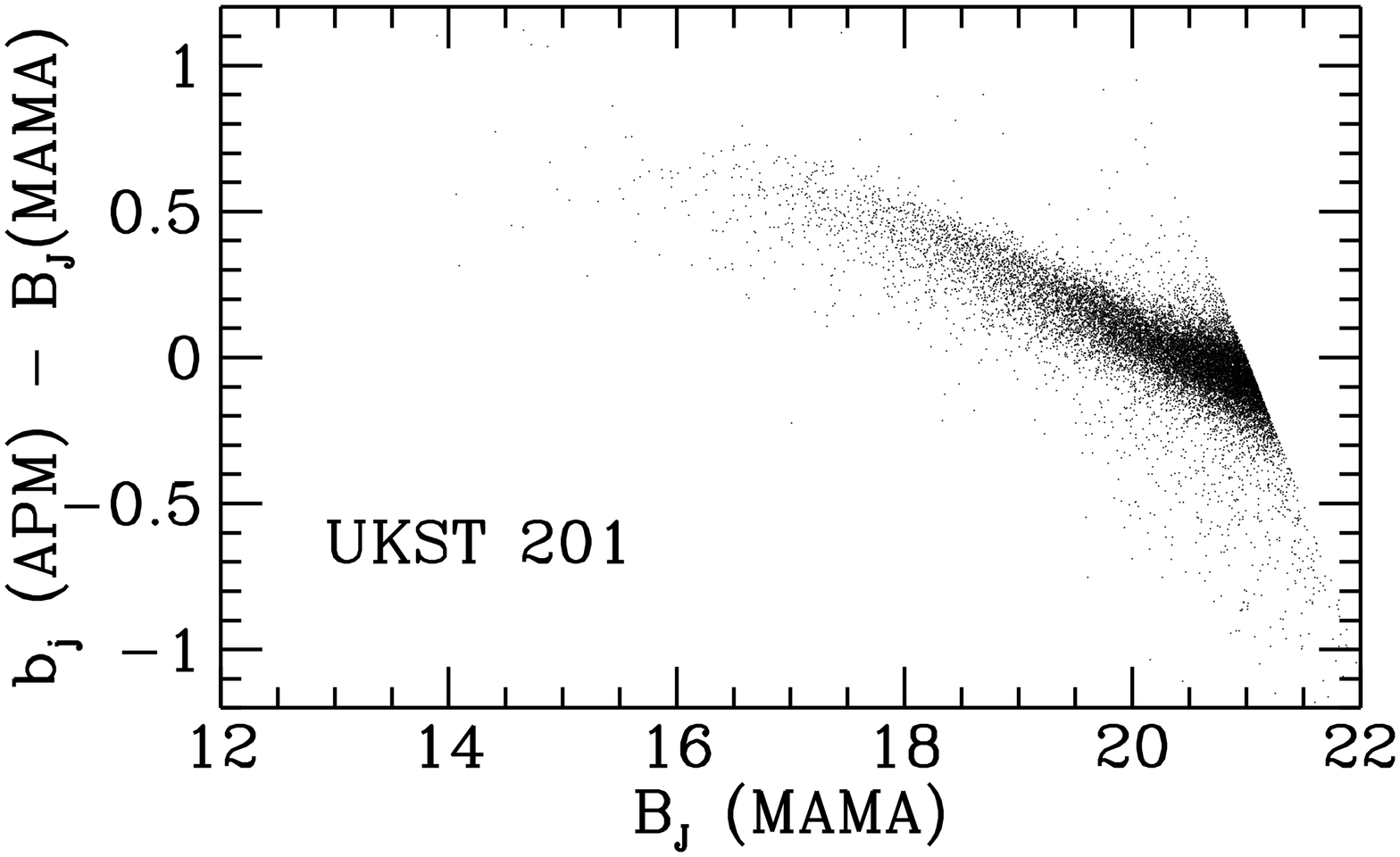,width=8.8cm,angle=-90}}}
 
  \centerline{\hbox{\psfig{figure=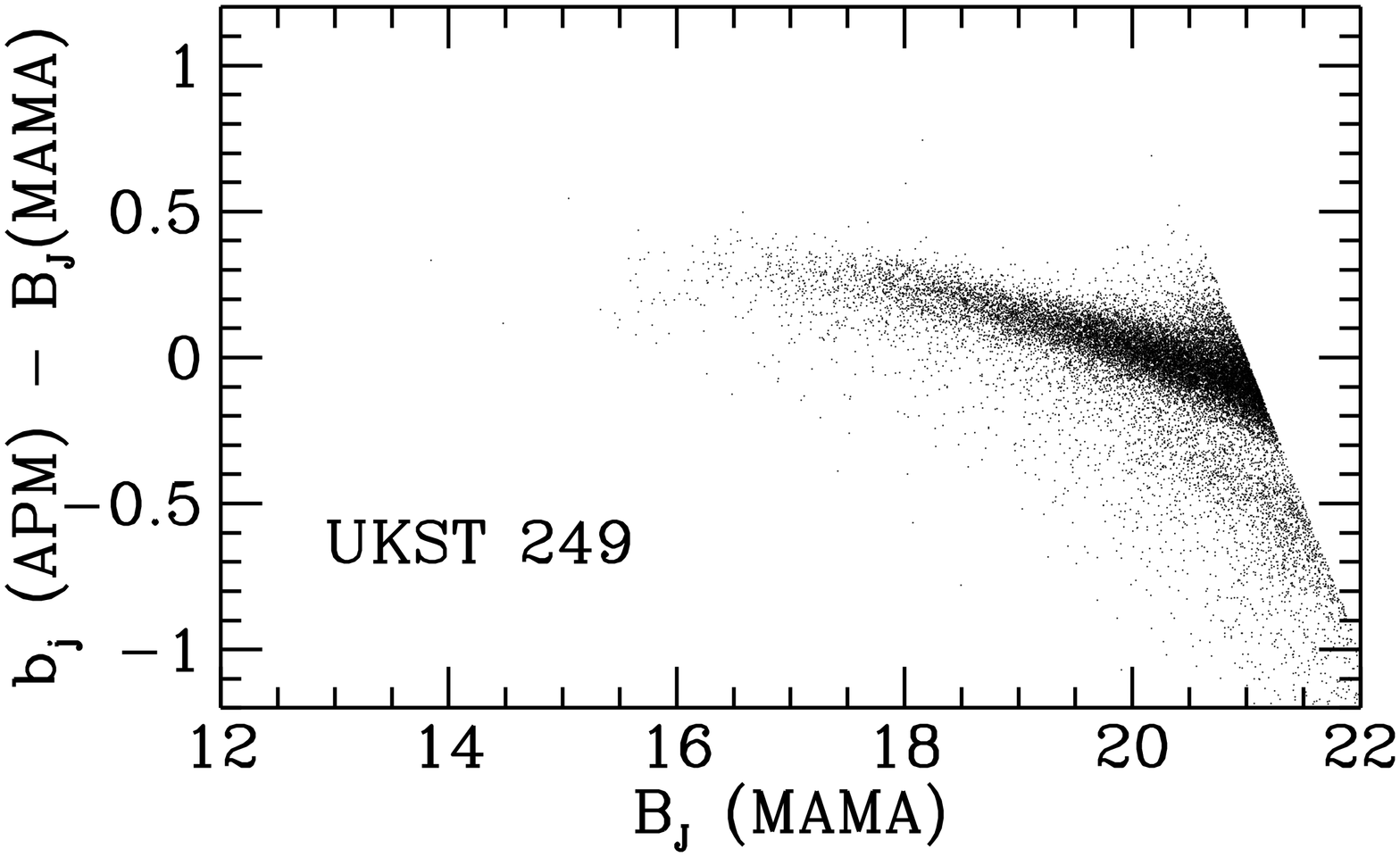,width=8.8cm,angle=-90}
                  \psfig{figure=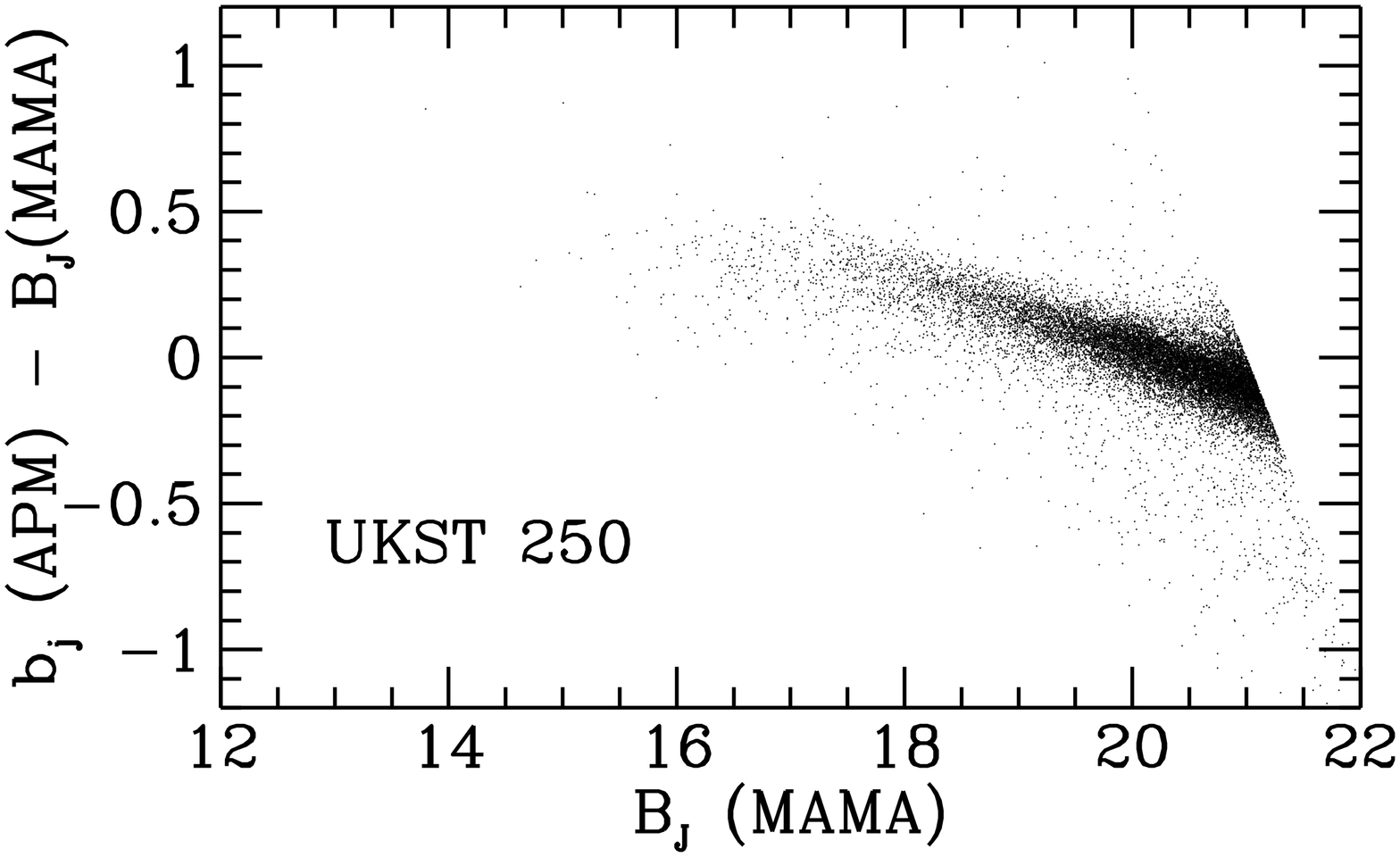,width=8.8cm,angle=-90}}}
  \caption[]{
       Comparison of galaxy magnitudes between the APM and our MAMA catalog on four UKST Schmidt plates. The tilted
       frontier on the right side of the plot is caused by the $\Bj =21$ limit in the APM subsample.}
   \label{fig:apmcompare}
   \end{figure*}

Figure \ref{fig:apmcompare} shows the difference between the APM catalog and our magnitudes for each of these four fields
\footnote{The version of the APM catalog shown here uses plate-to-plate matching based only on the faintest magnitude
bin ($19.5< b_j< 20.5$). The actual version used for statistical studies, limited to $b_j>17$, uses matching in
three magnitude ranges (G.~Dalton, private communication). However, a comparison between the two versions for our zone
shows an average difference (unmatched$-$matched) peaking to only $\approx +0.06$~mag. at $b_j = 17$.}.
Although both magnitudes are in agreement within $\approx 0.1$~mag. at the faint end ($\Bj\ge20$), one can immediately notice
a systematic difference at brighter magnitudes, reaching 0.3 to 0.5~mag. at $\Bj\approx 16-17$. To find out whether this was
due to our calibration or was intrinsic to the APM data, we cross-identified our list of southern standard galaxies with the APM
catalog to produce a plot similar to  Fig. \ref{fig:magdif}. Figure \ref{fig:apmstandcomp} shows indeed that there seems to be
some problem in the magnitude scale of this APM galaxy sample. The magnitude overestimate starts to rise at
$\Bj\approx 19.5$ and increases until $\Bj\approx 17$ where it culminates at about 0.4 mag. Brighter than this point, it is
unclear whether it stays at the same level or decreases again, but at $\Bj\approx 15-16$ the offset is still at least about
0.2 mag. Of course these estimates are in principle only valid for the small subsample of the APM catalog considered here,
and might not be applicable to the full catalog. But one can see that compensating for this average trend in the magnitude
scale would bring the total APM counts in very good agreement with ours.

\section{Comparison with models}
\label{par:modcomp}
\subsection{Numbers counts and the normalization of the ``local'' luminosity function}
\label{par:lf}
A recurring problem with previous counts has been the normalisation of models at bright magnitudes.
Normalizing passively evolving models at $\Bj=17$ leads to an apparent excess of already 50 to 100\%
at $\Bj \approx 20-21$. This apparent phenomenon is hard to reconcile with the redshift distributions
to $\Bj \la 23-24$ (Glazebrook et~al. \cite{glazebrook:al}, and references therein), which suggest a higher
normalisation of the field luminosity function.
As our data ``push'' the bright end of galaxy counts to higher values, we can now test how {\em local} determinations
of the luminosity function are able to match the counts over the domain $16 \le \Bj \le 21$. 

We therefore constructed a very simple non-evolving model of galaxy counts, intended to be valid at least to $\Bj\la 20$.
K-corrections were computed by integrating the spectral energy distributions from Pence (\cite{pence})
through the $b_j$ and $r_F$ photographic passbands given by Couch \& Newell (\cite{couch:newell}),
which are very close to our blue and red passbands.
The morphological mix of galaxy types was taken from Shanks et~al. (\cite{shanks:al}).
This model does not include luminosity or number evolution as in deeper count models
(e.g. Guiderdoni \& Rocca-Volmerange \cite{guiderdoni:rocca}, McLeod \& Riecke \cite{mcleod:rieke}),
but is sufficient for our normalisation purpose.
We chose an Einstein-de Sitter universe for simplicity,
although at this depth ($z \la 0.2$) the counts prove to be quite insensitive to (reasonable) cosmological parameters
(e.g. Yoshii \& Takahara \cite{yoshii:takahara}). 

\begin{figure}[htbp]
  \centerline{\psfig{figure=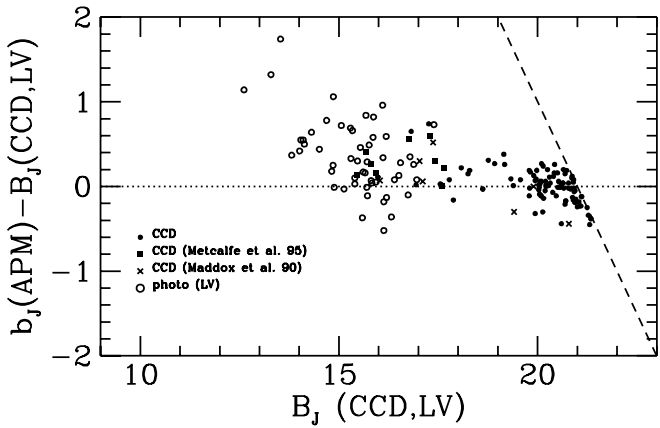,width=8.8cm,angle=-90}}
  \caption[]{
            Same as Fig. \ref{fig:magdif}, but with the APM catalog. The dashed line marks the magnitude limit of the APM
            subsample.}
   \label{fig:apmstandcomp}
   \end{figure}

\begin{figure}[htbp]
  \centerline{\psfig{figure=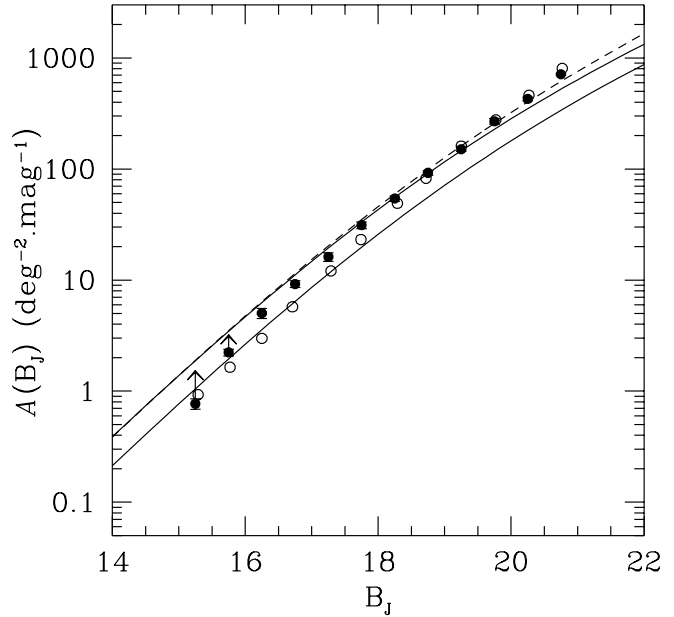,width=8.8cm}}
  \caption[]{
    Comparison of our $\Bj$ galaxy counts (filled circles) and those of the APM survey (open circles) with the simple models
    described in the text. Errorbars are $1 \sigma$ uncertainties deduced from the field-to-field scatter within
    each bin (assuming galaxy density is uncorrelated from one plate to another).
    Arrows indicate points affected by incompleteness. The 3 curves correspond to different
    luminosity functions used in the no-evolution model: Efstathiou \etal \cite{efstathiou:al} (upper solid line),
    Loveday \etal \cite{loveday:al} (lower solid line), and Shanks \cite{shanks} as normalized by Metcalfe \etal
    \cite{metcalfe:ala} (dashed line).}
   \label{fig:countmodcomp}
   \end{figure}

Figure \ref{fig:countmodcomp} shows our data compared to simple galaxy counts models in the blue photographic
passband without evolution as described above, using two different luminosity functions (LF). The first one is from
Efstathiou et~al. (\cite{efstathiou:al}, hereafter EEP), from a compilation of results obtained on several redshift
surveys, and the second one, more recent, is from Loveday et~al. (\cite{loveday:al}, hereafter LPEM) from a sparse
redshift survey of APM galaxies. Schechter parameters are given for both in Table \ref{tab:schechparams}.
A third, colour dependant, LF from Shanks (\cite{shanks}), with the normalisation adopted by Metcalfe \etal
(\cite{metcalfe:ala}) is also shown for comparison.
As can be seen, the EEP luminosity function fits
the data much better than the one from LPEM, which are however, as expected, in good agreement with the APM counts for
$\Bj\le17$. In fact, most of the discrepancy between the two luminosity functions lies in the value of $M_{\tiny \Bj}^*$,
with a difference close to 0.4~mag, i.e. about the same one gets when comparing APM magnitudes to those of standard
galaxies at $\Bj\approx 17$ (Fig. \ref{fig:apmstandcomp}).
The best agreement is found with the Shanks LF model. This is not a surprise, as the two other models do not
distinguish the luminosity functions of early and late-types, which are indeed
quite different; and this shows up when K-corrections become important.

\begin{table*}[htbp]
\label{tab:schechparams}
\caption{Global Schechter parameters of local luminosity functions ($H_0 = 100 {\rm km.s}^{-1}{\rm Mpc}^{-1}$)}
\begin{tabular}{lccc}
     \hline
     & $\alpha$ & $M_{\tiny \Bj}^*$ & $\phi^*$ (${\rm Mpc}^{-3}$) \\
     \hline
     \hline
Efstathiou et~al. (\cite{efstathiou:al}) & $-1.07 \pm 0.05$ & $-19.97^a \pm 0.10$ & $(1.56\pm 0.34)\times 10^{-2}$ \\
Loveday et~al. (\cite{loveday:al})       & $-0.97 \pm 0.15$ & $-19.50   \pm 0.13$ & $(1.40\pm 0.17)\times 10^{-2}$ \\
     \hline
\end{tabular}\\
{\small $^a$ Converted from ${\rm B}_{\rm T}$ using Eq. (4.1d) from Efstathiou et~al.}
\end{table*}

Figure \ref{fig:countmodcomp} indicates that the normalisation of the Shanks LF adopted by Metcalfe \etal (\cite{metcalfe:ala})
is in excellent agreement with our own counts, but the latter are still slightly too steep on their bright side compared
to any non-evolving model. Can this be interpreted as evolution? One can hardly conclude on the basis of the data
presented here, as a 10\% misclassification of galaxies added to a 0.1 mag.
systematic error in measured fluxes at $\Bj \approx 17$ would be enough to produce this effect.

\subsection{Galaxy colours}
\label{par:colours}
\begin{figure*}[htbp]
  \centerline{\psfig{figure=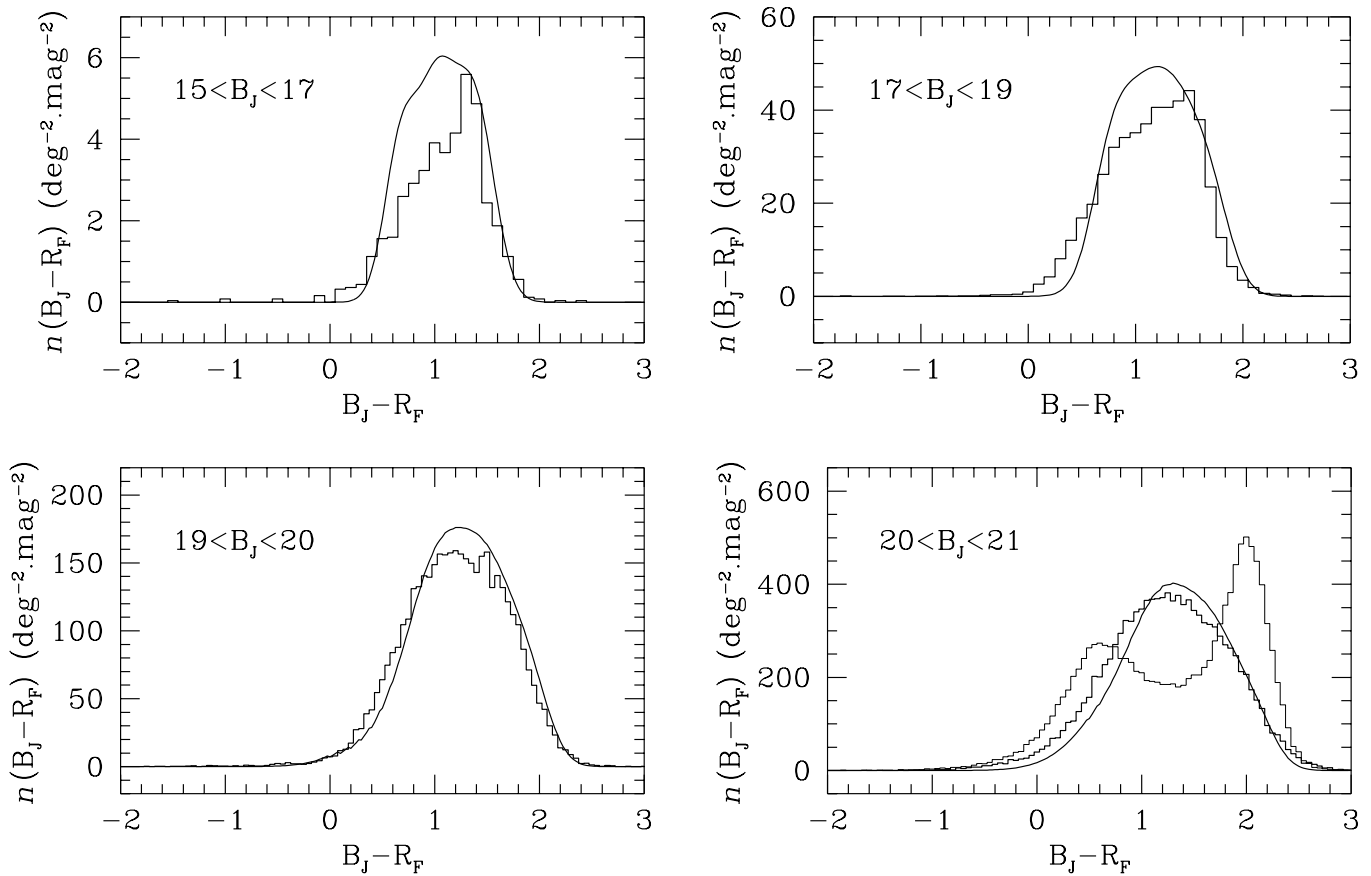,width=18cm,angle=-90}}
  \caption[]{
            Colour histograms of galaxies compared with the no-evolution model (curves) described in the text, in 4 magnitude intervals.
            The colour distribution of stars (thin histogram) is plotted for comparison in the faintest subsample.}
   \label{fig:galcol}
   \end{figure*}

Galaxy colour distributions for the total catalog (including the field 16\fh+42{\degr}) are presented in
Fig. \ref{fig:galcol}. They are in good agreement with the compilation presented by Koo \& Kron (\cite{koo:kron}),
if one makes allowance for the passband differences between our system and theirs.

We reemployed the Shanks LF model with no evolution described above to model the observed colour distribution.
Rest-frame galaxy colours per Hubble type were taken from Metcalfe \etal (\cite{metcalfe:ala}) and converted from their CCD
photometric passbands to our photographic system with $(\Bj-\Rf) = 0.91(\B-\R)_{\mbox{\tiny CCD}}$. Ingredients of the LF
are gathered in Table \ref{tab:colparam}. Photometric errors were included in the model, adopting an error distribution
with magnitude determined in the overlap between catalogs (see Bertin \cite{bertinb}). Photometric errors do not only
smear out details in the observed colour distribution, they also distort the wings of the distribution for the dimmest objects
(that is, when the error strongly evolves with colour). As our red plates are here significantly less sensitive than the blue
ones, the two lower graphs in Fig. \ref{fig:galcol} exhibit a shallower tail on their blue side than on their red side.

Observed and predicted distributions prove to be in very good agreement, except in the brightest subsample, where we expect
some problems with the saturation and the robustness of star/galaxy separation. One can notice a small ($\approx 0.1$~mag)
systematic colour offset between the two, which might be imputed to uncertainties in our photometric modelling.
But no obvious evolution of colour with magnitude is detected, within the uncertainties, up to $\Bj\la21$.

\begin{table}[htbp]
\label{tab:colparam}
\caption{Luminosity function parameters for the colour distribution model ($H_0 = 100 {\rm km.s}^{-1}{\rm Mpc}^{-1}$)}
\begin{tabular}{lcccc}
     \hline
     & $\alpha$ & $M_{\tiny \Bj}^*$ & $\phi^*$ (${\rm Mpc}^{-3}$) & $\langle\Bj-\Rf\rangle$ \\
     \hline
     \hline
     E/S0 & -0.7 & -19.6 & $1.2\times10^{-3}$ & 1.32 \\
     Sab  & -0.7 & -19.6 & $3.9\times10^{-3}$ & 1.15 \\
     Sbc  & -1.1 & -19.9 & $6.9\times10^{-3}$ & 0.96 \\
     Scd  & -1.5 & -20.1 & $2.7\times10^{-3}$ & 0.70 \\
     Sdm  & -1.5 & -20.1 & $1.7\times10^{-3}$ & 0.62 \\
     \hline
\end{tabular}\\
\end{table}

\section{Discussion}
\label{par:discuss}

\subsection{A magnitude scale error in the APM data?}
The apparent rapid increase in number counts of galaxies in the magnitude range $16 \la \Bj \la 19$ reported by
Maddox et~al. (\cite{maddox:ald}) is thus absent from our counts (or those of Weir et~al. \cite{weir:al}), although
our slope of $dN/dm$ is still slightly higher than what predict no-evolution models at $\Bj\la17$.
As the comparison with the APM data seems to indicate, this might be essentially due to a difference in magnitude
scale of $\approx $~0.1-0.2~mag./magnitude. Serious suspicions about systematic residual errors in the APM magnitudes
at $\Bj \approx 17-18$ have recently been raised by Metcalfe et~al. (\cite{metcalfe:alb}). If our data are correct,
they support and even strengthen this hypothesis. Other arguments also consolidate this idea.

First, our data (Fig. \ref{fig:galcol}), as well
as previous ones (see Koo \& Kron \cite{koo:kron}), do not present evidence for evolution of galaxy
colours over the range $17\le\Bj\le21$. and rule out strong luminosity evolution scenarios in the visible
(McLeod \& Rieke \cite{mcleod:rieke}) for bright field galaxies, out to $z \approx 0.2$.

One might also think of a ``giant local void'' that would affect the bright end of the counts. Although there are indications
of large fluctuations in projected galaxy density at the level of a Schmidt field (Fig. \ref{fig:multicounts},
Picard \cite{picardb}) for $\Bj \la 17$, or over the full sky for $\Bj \la 12$ (Paturel et~al. \cite{paturel:alb}), the analysis
of a large spectroscopic subsample (Loveday et~al. \cite{loveday:al}) seems to exclude this possiblity in the APM survey.

Statistics of CCD calibrations (Maddox et~al. \cite{maddox:ald}, Loveday et~al. \cite{loveday:al}) done on the APM
magnitudes do not reveal any important trend like the one in Fig. \ref{fig:apmcompare} and \ref{fig:apmstandcomp}.
This is quite puzzling, and one might wonder if the effects seen here are not purely local and compensated somewhere else
in the APM survey. If this were true, correlated systematic magnitude errors
of $\approx 0.4$~mag., over such large areas (100 sq.~deg. in our case) would artificially boost the two-point correlation
function on scales $\ga 5{\degr}$ in $\Bj \la 18$ magnitude bins. This does not appear prominently in Fig. 2 of
Maddox \etal (\cite{maddox:ala}). Another clue is that the trend found by Metcalfe et~al. (\cite{metcalfe:alb}) concerns
the whole area of the APM catalog, and therefore supports the hypothesis of a global magnitude scale error (which
does not exclude large scale variations of the zero-point).

\subsection{Uncertainties in the ``local'' field luminosity function}
To  partially bypass uncertainties in the optimum Schechter parameters of the local luminosity function for field galaxies,
count models are generally ``renormalized'' by adjusting the Schechter density parameter $\phi^*$ to counts at some bright
magnitude. However, the $M^*$ parameter itself is directly affected by any systematic error in the magnitude zero-point of
a redshift survey. Such errors are known to exist at bright magnitudes in photographic catalogs
(Metcalfe \etal \cite{metcalfe:alc}, Yasuda et~al. \cite{yasuda:al}) on which are
essentially based all the determinations of the local field luminosity function. In fact, decreasing the $M^*$ found
by LPEM by about 0.4~mag., like our photometric comparison with the APM catalog suggests, brings their luminosity function
in good agreement with both the one from EEP and our bright galaxy counts (Fig. \ref{fig:countmodcomp}),
without having to increase $\phi^*$.
One should also note that such a``bright'' normalisation in $M^*$ of the luminosity function removes any significant luminosity
evolution\footnote{The related case of evolution in the far-infrared as seen by IRAS will be addressed in a forthcoming paper
(Bertin et~al., in preparation).} of luminous galaxies in the blue band to $z\approx 0.2$
(Lonsdale \& Chokshi \cite{lonsdale:chokshi}).

\section{Summary and Conclusions}
We have presented a new photometric survey of bright galaxies with $15<\Bj<21$ and $14<\Rf<19.5$. Comparison with
our own CCD galaxy standards, as well as others from the literature, allowed us to estimate systematic errors in the photometry
to be $\la 0.1$~mag over the whole magnitude range of the survey. Our data reveal no evidence of strong
galaxy evolution as had been reported by Maddox et~al. (\cite{maddox:ald}) with the APM survey, although the slope
of our number counts is still somewhat higher than what is predicted by no-evolution models brightward $\Bj\approx 17$.
A comparison with the APM data
and other arguments suggest that the APM catalog is affected by a large magnitude scale error, underestimating by about
0.4 mag the flux
of galaxies with $\Bj\la 17$. This would then justify the choice of a ``high'' normalization of the field luminosity
function (concerning $M^*$ as well as $\phi^*$) already used by several authors to improve the fit of models at $\Bj>19$
(e.g. Metcalfe et~al. \cite{metcalfe:ala},\cite{metcalfe:alb}, Glazebrook et~al. \cite{glazebrook:al}).

But, given the difficulties inherent to photographic photometry, it would be good to confirm our analysis with linear
detectors in order to conclude definitively.
Forthcoming digital large scale surveys like the SDSS (e.g. Kent \cite{kent}) should provide this possibility.

\begin{acknowledgements}
We thank the team from CERGA, especially D.Albanese and C.Pollas for providing the northern Schmidt plates;
the MAMA team, especially R.~Chesnel and O.~Moreau for their high-quality scans; P.Fouqu\'e for enlightning
discussions about the local universe and galaxy photometric systems; J.~Loveday and G.~Dalton for providing
the APM comparison subsamples; and our referee for useful suggestions.

\end{acknowledgements}

\end{document}